\documentclass[12pt,preprint]{aastex}

\newcommand{\Hipp}{{\it Hipparcos}}        

\newcommand{\HST}{{\it HST}}

\newcommand{\Teff}{T_{\rm eff}}
\newcommand{\Mjup}{M_{\rm Jup}}

\newcommand{\kms}{{\>\rm km\>s^{-1}}}

\shorttitle{HST Astrometry of Sirius}
\shortauthors{Bond et al.}

\begin{document}

\title{The Sirius System and its Astrophysical Puzzles: \\
{\em Hubble Space Telescope\/} and Ground-Based Astrometry\altaffilmark{1} 
}

\author{
Howard E. Bond,\altaffilmark{2,3}
Gail H. Schaefer,\altaffilmark{4}    
Ronald L. Gilliland,\altaffilmark{3,5} 
Jay B. Holberg,\altaffilmark{6} 	
Brian D. Mason,\altaffilmark{7} 	
Irving W. Lindenblad,\altaffilmark{7,8} 	
Miranda Seitz-McLeese,\altaffilmark{7,9} 
W. David Arnett,\altaffilmark{10}  
Pierre Demarque,\altaffilmark{11}   
Federico Spada,\altaffilmark{12}  
Patrick A. Young,\altaffilmark{13}  
Martin A. Barstow,\altaffilmark{14}
Matthew R. Burleigh,\altaffilmark{14}
and
Donald Gudehus\altaffilmark{15}   	
}

\altaffiltext{1} 
{Based in part on observations with the NASA/ESA {\it Hubble Space Telescope\/}
obtained at the Space Telescope Science Institute, and from the Mikulski Archive
for Space Telescopes at STScI, which are operated by the Association of
Universities for Research in Astronomy, Inc., under NASA contract NAS5-26555}

\altaffiltext{2}
{Department of Astronomy \& Astrophysics, Pennsylvania State
University, University Park, PA 16802, USA; heb11@psu.edu}

\altaffiltext{3}
{Space Telescope Science Institute, 
3700 San Martin Dr.,
Baltimore, MD 21218, USA}

\altaffiltext{4}
{The CHARA Array of Georgia State University, Mount Wilson Observatory,
Mount Wilson, CA 91023, USA}

\altaffiltext{5}
{Center for Exoplanets and Habitable Worlds,
Department of Astronomy \& Astrophysics, Pennsylvania State
University, University Park, PA 16802, USA}

\altaffiltext{6} 
{Lunar \& Planetary Laboratory, University of Arizona, 1541 E. University
Blvd., Tucson, AZ 85721, USA}

\altaffiltext{7}
{U.S. Naval Observatory, 3450 Massachusetts Ave., Washington, DC 20392, USA}

\altaffiltext{8}
{Deceased 2011 November 11}

\altaffiltext{9}
{USNO SEAP intern, 2009-10; current address: Institute for Defense Analyses,
4850 Mark Center Dr., Alexandria, VA 22311, USA}

\altaffiltext{10}
{Steward Observatory, 933 N. Cherry Ave., University of Arizona, Tucson, AZ
85721, USA}

\altaffiltext{11}
{Department of Astronomy, Yale University, Box 208101, New Haven, CT 06520, USA}

\altaffiltext{12}
{Leibniz-Institut f\"ur Astrophysik Potsdam, An der Sternwarte 16, D-14482,
Potsdam, Germany}

\altaffiltext{13}
{School of Earth \& Space Exploration, Arizona State University, Tempe, AZ
85287, USA}

\altaffiltext{14}
{Department of Physics \& Astronomy, University of Leicester, Leicester LE1
7RH, UK}

\altaffiltext{15}
{Department of Physics \& Astronomy, Georgia State University, Atlanta, GA
30303, USA}

\begin{abstract}

Sirius, the seventh-nearest stellar system, is a visual binary containing
the metallic-line A1~V star Sirius~A, brightest star in the sky, orbited in a
50.13-year period by Sirius~B, the brightest and nearest white dwarf (WD)\null.

Using images obtained over nearly two decades with the {\it Hubble Space
Telescope\/} (\HST), along with photographic observations covering almost 20
years, and nearly 2300 historical measurements dating back to the 19th century,
we determine precise orbital elements for the visual binary. Combined with the
parallax and the motion of the A component, these elements yield dynamical
masses of $2.063\pm0.023\,M_\odot$ and $1.018\pm0.011\,M_\odot$ for Sirius~A and
B, respectively.

Our precise \HST\/ astrometry rules out third bodies orbiting either star in the
system, down to masses of $\sim$15--$25\,M_{\rm Jup}$. 

The location of Sirius~B in the H-R diagram is in excellent agreement
with theoretical cooling tracks for WDs of its dynamical mass, and implies a
cooling age of $\sim$126~Myr. The position of Sirius~B in the mass-radius plane
is also consistent with WD theory, assuming a carbon-oxygen core. Including the
pre-WD evolutionary timescale of the assumed progenitor, the total age of
Sirius~B is about $228\pm10$~Myr.

We calculated evolutionary tracks for stars with the dynamical mass of Sirius~A,
using two independent codes. We find it necessary to assume a slightly sub-solar
metallicity, of about $0.85\,Z_\odot$, to fit its location in the
luminosity-radius plane. The age of Sirius~A based on these models is about
237--247~Myr, with uncertainties of $\pm$15~Myr, consistent with that of the WD
companion.

We discuss astrophysical puzzles presented by the Sirius system, including the
probability that the two stars must have interacted in the past, even though
there is no direct evidence for this, and the orbital eccentricity remains high.

\end{abstract}


\section{The Sirius Binary System}

Sirius ($\alpha$~Canis Majoris), brightest star in the sky, belongs to the
seventh-nearest stellar system, at a distance of only 2.6~pc. Periodic
astrometric perturbations of its proper motion---and of the other ``Dog Star,''
Procyon ($\alpha$~Canis Minoris)---were discovered by Bessel (1844), who
recognized that they must be caused by dark satellites. The faint companion,
Sirius~B, was first seen visually in 1862 by Alvan G. Clark and his father, as
reported and confirmed by Bond (1862a). Sirius~B was noted moreover to lie at a
position angle consistent with it being the perturbing body inferred by Bessel,
but to be so faint compared to its gravitational influence that Bond (1862b)
stated that it must be ``only feebly self-luminous.''  

The spectrum of Sirius~B  was photographed by Adams (1915), and found to be
remarkably similar to that of the main-sequence early A-type primary
Sirius~A\null. Along with the earlier discovery of an A-type spectrum for the
intrinsically faint star o$^2$~Eridani~B, this finding (as famously recounted
three decades later by Russell 1944) established the existence of a new class of
stars with low luminosities but relatively early spectral types. The term
``white dwarf'' was coined for these objects by Luyten (1922). Sirius~B is the
nearest and brightest white dwarf (WD), cataloged as WD~0642$-$166. It is a hot
WD with a pure-hydrogen photosphere and a spectral type of DA2. The classical
spectral type of Sirius~A is A1~V (Morgan et al.\ 1953), but high-resolution
spectra reveal surface overabundances of heavy elements by factors of as much as
$\sim$10 to 100 relative to solar (e.g., Cowley et al.\ 2016 and references
therein). Thus Sirius~A is generally considered to be a hot metallic-line (Am)
star. Its rotational velocity is very low compared to typical A-type stars (Gray
2014 gives $v\sin i=16.7\kms$), which favors the action of levitative processes
in the outer layers.

It was already apparent by the mid-19th century, based on the astrometric
perturbation, that the orbital period of the Sirius system is close to 50~years
(e.g., Auwers 1864). Compared to the notoriously difficult Procyon~B, Sirius~B
is somewhat easier to detect visually or photographically, except around the
time of closest separation from A; however, the difference in visual brightness
of about ten magnitudes leads to comparatively large uncertainties, both random
and systematic, in measurements of the separation and position angle (PA)\null.
When slightly more than one revolution of the system had been observed, Aitken
(1918) analyzed the available observations of Sirius~B (all of them made
visually with micrometers) and determined orbital elements. Later, based on
measurements covering nearly two revolutions, and now including photographic
observations, a new orbital solution was published by van den Bos (1960). It
yielded dynamical masses for Sirius~A and B of 2.15 and $1.05\,M_\odot$,
respectively. Gatewood \& Gatewood (1978, hereafter GG78), employing
measurements of over 300 photographic plates obtained at the Yerkes and
Allegheny Observatories between 1917 and 1977, refined the orbital elements, but
found nearly the same masses, 2.14 and $1.05\,M_\odot$. Sirius~B is thus one of
the most massive known WDs, particularly among the nearby sample. Historical
details of these discoveries and subsequent developments have been recounted by
several authors, especially thoroughly by van de Kamp (1971), GG78, Hetherington
(1980), Holberg \& Wesemael (2007), Holberg (2007, 2009, 2010), Brosch (2008),
and Wesemael \& Racine (2008).

In contrast with ground-based observations, Sirius~B is easily resolved in
appropriately exposed images obtained with the {\it Hubble Space Telescope\/}
(\HST\/)\null. Because the binary is so nearby, and the atmospheric parameters
of both stars are well known, the Sirius system offers the possibility of
fundamental constraints on stellar physics for both main-sequence stars and
massive WDs. With this in mind, in 2001 our team began a program of regular
\HST\/ imaging and astrometry of the binary. Our aims were to obtain dynamical
masses of both stars with the highest possible precision, and an accuracy
limited only by the absolute parallax of the system. Moreover, precise relative
astrometry of the binary would place limits on---or could detect---the presence
of third bodies in the system, down to substellar masses. 

Our project began with imaging using the Wide Field Planetary Camera~2 (WFPC2)
in 2001 October, and we observed the Sirius system with \HST\/ at a total of 10
epochs until 2008 January. We then continued the program with the Wide Field
Camera~3 (WFC3), following its installation in place of WFPC2 during the 2009
\HST\/ Servicing Mission. We obtained WFC3 frames at five epochs between 2010
September and 2016 August. In addition, the \HST\/ archive contains WFPC2
observations at two epochs in 1997, making a grand total of 17 epochs between
1997 and 2016, a time interval covering almost 40\% of the orbital period. Since
the binary astrometry is no longer the dominant constraint on the dynamical mass
determinations, we have now concluded the \HST\/ imaging program, and we present
our final results here.

Supplementing our highly precise \HST\/ astrometry are nearly 2300 published
ground-based observations of Sirius, obtained between 1862 and 2016. We made a
literature search and critical analysis of these data, and include them (with
appropriate vetting and statistical weights) in our determination of the orbital
elements of the binary. We also present 66 previously unpublished photographic
measurements made at the U.S. Naval Observatory (USNO) between 1970 and 1984. We
then derive precise dynamical masses for both components of the binary, discuss
the astrophysical implications, and place limits on the presence of third bodies
in the system. Our study closely parallels a similar presentation of \HST\/ and
ground-based astrometry of the Procyon system (Bond et al.\ 2015, hereafter
B15).

\section{{\em HST\/} Observations}

The visual magnitude of Sirius~A is $V=-1.47$ (Johnson \& Morgan 1953).
Sirius~B, at $V=8.44$ (Holberg et al.\ 1998), is fainter by a factor of 9,200.
Astrometry of this binary, even with \HST, therefore presents two observational
challenges: the extreme brightness of the primary star, and the extremely large
flux ratio. There is no combination of a narrow-bandpass filter and short
exposure time with \HST\/ using either WFPC2 or WFC3 that would not result in
saturated pixels in the image of Sirius~A\null. We thus adopted a strategy of
obtaining frames with exposure times long enough to show Sirius~B with good
signal-to-noise ratio (SNR), in which Sirius~A was allowed to be grossly
overexposed. Unsaturated features in the outer regions of its point-spread
function (PSF), principally the diffraction spikes, would be used to determine
its centroid location. (We successfully employed this same approach for our WFC3
imaging of Procyon, as described in detail in B15.)

For the WFPC2 camera, we considered several possible bandpasses, but selected
the one at the longest available wavelength, the F1042M filter centered near
$1.0\,\mu$m. This filter had the advantages of (1)~a PSF with a well-defined
``triple'' structure of the diffraction spikes (due to the first Airy ring),
(2)~a relatively low system throughput, and (3)~availability of a substantial
number of archival images in this filter, including archival frames of Sirius
itself. The main disadvantage of F1042M is that the flux ratio between A and B
is even higher than in the visual, about a factor of 18,000, due to the hotter
temperature of B\null.

We placed Sirius near the center of the Planetary Camera (PC) chip of WFPC2,
providing $800\times800$ pixel images with a plate scale of $0\farcs0454 \, \rm
pixel^{-1}$. We specified telescope roll angles such that Sirius~B would not lie
near the diffraction spikes or charge bleeding of the bright component, and
obtained images at several different dither locations during each visit. We used
a range of exposure times from 4 to 60~s, along with a few very short exposures
(0.11~s, the shortest possible with WFPC2), which we ended up not using in our
final analysis.\footnote{There are also limited archival \HST\/ observations of
Sirius obtained with WF/PC-1, NICMOS, and STIS, and with other WFPC2 filters
than the ones we used, but we judged these unlikely to contribute additional
useful astrometric data.} 

For the WFC3 observations, we chose the longest-wavelength narrow-band filter
available in the UVIS channel, F953N, with a similar strategy of dithered images
of the binary in which Sirius~A is overexposed. We used a $1024\times1024$ pixel
subarray (in order to reduce data volume and allow more frames to be taken
during each \HST\/ visit), with a pixel scale of $0\farcs0396 \, \rm
pixel^{-1}$, and exposure times of 6 and 12~s.

An observing log for the WFPC2 and WFC3 data is presented in Table~1. The first
two lines give details for archival WFPC2 visits in 1997, which had been
obtained as part of an (unsuccessful) search for new faint companions of nearby
stars (Schroeder et al.\ 2000). We included these frames in our astrometric
study.

\section{{\em HST} Astrometric Analysis}

For the measurements of separation and PA for the Sirius system we
have two sets of \HST\/ data. These are (1)~WFPC2/PC frames in the F1042M
filter; and (2)~WFC3/UVIS frames in the F953N filter. 

\subsection{WFPC2 Images in F1042M}

Figure~1 illustrates a typical WFPC2 frame, containing a severely overexposed
image of the primary star and a well-exposed image of Sirius~B lying to the
lower left. Our WFC3 frames have a similar appearance.

For the astrometric measurements of these frames, we followed an essentially
identical procedure to that described in B15 for our analysis of overexposed
WFPC2 F1042M images of Procyon; thus we do not repeat all of the details here
but only give a brief outline. In particular, in order to build over-sampled
representations of the F1042M PSF for their study of Procyon, B15 included and
discussed an analysis of all the available WFPC2 data on Sirius. As Figure~1
shows, the diffraction spikes exhibit quasi-periodic variations in intensity as
a function of distance from the center. B15 found that this structure varies
from epoch to epoch (probably because of a dependence on the exact location of A
within the field of view), making it difficult to construct a usable
over-sampled PSF for the unsaturated outer regions of the images\null. Instead,
we used a procedure of fitting straight lines to the unsaturated portions of the
diffraction spikes, and defining their intersection point to be the centroid of
Sirius~A\null. For the unsaturated images of Sirius~B, centroids were determined
from a more conventional technique of PSF fitting, again as discussed in detail
in B15. In order to calibrate the systematic offset in the centers using the two
different techniques, we obtained unsaturated F1042M images of the star
109~Virginis, an A0~V star with a color very similar to that of Sirius, but
sufficiently faint that both unsaturated and saturated images could be compared
directly. The resulting offsets were applied to the spike intersection points to
place them in the same system as the direct centroids of Sirius~B.  

\subsection{WFC3 Images in F953N}

A similar approach was used for the WFC3 observations of Sirius in F953N, which
is described in detail in B15 for the analysis of images of Procyon obtained in
the same filter. As for the WFPC2 data, we also analyzed and discussed all of
the Sirius data then available with WFC3 as direct support of the Procyon
results in B15. Again, we fitted straight lines to the diffraction spikes to
determine the centroid of A, and used PSF fitting to find the location of B in
each image. The offset between the two methods was determined from unsaturated
and saturated calibration frames we obtained for the A3~V star HD~23886.
However, one difference from the WFPC2 frames is that the diffraction spikes in
F953N do appear to have a consistent appearance at all epochs. Thus for the WFC3
frames we could use an alternative method of developing a deep mean PSF for the
outer regions, based on a large number of frames of Sirius, Procyon, and
HD~23886. This allowed us to determine the positions of Sirius~A using PSF
fitting instead of the diffraction-spike intersection point. The positions of
Sirius~B were again found from a conventional unsaturated PSF-fitting method.
The systematic offset correction between the two PSF-based centroids was
determined from the unsaturated and saturated calibration observations of
HD~23886.   

\subsection{Astrometric Results from {\em HST}}

The process of converting the astrometry from the image plane to the absolute
J2000 frame is, again, described in B15. It is based on an adopted plate scale,
and the known orientation on the sky of the image $y$ axis. Our final \HST\/
astrometric results are given in Table~2. For WFC3 we present the results from
both methods---the diffraction-spike intersection and the PSF fit---as well as
the weighted means of the two. Note that the PAs are referred to the equator of
J2000, not to the equator of the observation epoch that is the usual practice
for ground-based measurements.

\section{Ground-based Measurements}

As in the case of Procyon (see B15), the determination of orbital elements for
Sirius is improved through inclusion of historical observations---because of
their much longer time coverage than provided by the \HST\/ data, and because of
measures obtained at orbital phases not observed by \HST\null. In the first
subsection below we present a set of previously unpublished historical
photographic observations. In the second subsection we refer to our new critical
compilation of all published ground-based astrometric measurements of Sirius. 

\subsection{USNO Photography, 1965--1984}

A long-term program of photographic astrometry of the Sirius system was started
in 1965 by I.W.L.,\footnote{Irving W. Lindenblad passed away on 2011
November~11. An obituary is available at \tt
https://aas.org/obituaries/irving-w-lindenblad-1929-2011} using the 26-inch
refractor of the USNO in Washington, DC. These observations used a hexagonal
aperture mask, which causes intensity to drop off rapidly in certain directions
around a bright star, while producing six bright ``spikes'' at other PAs (e.g.,
Aitken 1935, p.~60; van Albada 1962). Proper mask orientation allows close
companions to be detected that would not otherwise be resolved easily. Along
with the hexagonal mask, an objective grating with extremely fine, evenly spaced
wires was added. This produced good first- and second-order images of the
primary star, which could be measured and used to locate its centroid. Figure~2
shows a digitized version of a typical photographic plate from this series of
observations.

Trailed exposures were also taken on the same plates, to define the east-west 
direction. Details of the observations, and astrometric measurements of the 
separation and PA on 56 nights between 1965 and 1969, were published
by Lindenblad (1970). A later paper (Lindenblad 1973) presented additional
measurements between 1969 and 1972. Subsequently this observing program was
continued until 1984, but unfortunately these plates had never been measured or
the results published. During this interval, about 160 usable photographic
observations were obtained, on 66 different nights.

Astrometry of these plates was carried out by M.S.-M. by digitizing them on the
StarScan (Zacharias et al.\ 2008) machine at USNO\null. A centroiding algorithm
was used for the images of Sirius~A and B, with the mean position of the two
first-order images of A defining its location. A plate scale of $20\farcs8476
\,\rm mm^{-1}$ was adopted.

These measures were corrected for the ``Ross effect'' (see Lindenblad 1970),
whereby the blackened portions of photographic emulsions dry faster than the
non-blackened, causing nearby portions of the emulsion to contract
differentially. The correction was determined by using measures of the
second-order images. These images are far enough away from the bright primary
that they are essentially undisturbed by the Ross effect. In the absence of
contraction, the distance between the second-order images would be twice the
distance between the first-order images, and the departure from this relation
allows calculation of the correction (van Albada 1962, 1971).

Table~3 presents the results of these measurements. The PAs are for the equator
of the observation epoch.

\subsection{Critical Compilation of Historical Data, 1862--2016}

We have collected and critically examined all published measurements of the
Sirius visual binary of which we are aware (to which we add the new \HST\/ and
USNO data presented in this paper). These data are discussed in Appendix~A,
along with excerpts from the associated full electronic versions of the tables.

\section{Elements of the Relative Visual Orbit of Sirius B}

\subsection{Corrections to J2000}

The first step in our determination of orbital elements was to adjust all of the
measurements, both \HST\/ and ground-based, to a J2000 standard equator and
epoch. We used the formulations given by van den Bos (1964) in order to correct
for (1)~precession (except for the \HST\/ measures, which are already in the
J2000 frame), (2)~the change in direction to north due to proper motion, (3)~the
changing viewing angle of the three-dimensional orbit due to proper motion,
and (4)~the steadily decreasing distance of the system due to radial velocity
(RV)\null.

Except for precession, these corrections are small relative to the observational
uncertainties for the ground-based data, and are also small for the \HST\/ data
because their epochs are all so close to 2000.0. We verified our coding by
showing that it reproduces the values presented for Sirius by van den Bos
(1960), if we used his input parameters. For the actual corrections, we used the
parallax from \S6.1 below, the {\it Hipparcos\/} proper motion (van Leeuwen
2007), and for the space-motion correction an initial solution for the orbital
elements of the visual binary. The RV of the center of mass of the Sirius system
can be determined from the observed RVs of Sirius~A and the relative orbit.
There have been several such determinations, beginning with Campbell (1905), who
obtained $-7.4\,\kms$. Here we use our relative orbit and all available RV data
from 1903 to 1995 to determine a system velocity of  $-7.70\,\kms$. (Details of
the RV determination will be presented in a separate forthcoming spectroscopic
paper on Sirius~B\null.) To obtain the true radial component of the space motion
we must correct this for the gravitational redshift of Sirius~A, yielding
$-8.47\,\kms$.

In Appendix~A we tabulate both the input observed PAs and separations, and those
corrected to J2000.

\subsection{Orbital Solution}

We determined elements for the visual orbit via a seven-parameter fit to the
combined set of J2000-corrected \HST\/ and ground-based measurements. This fit
employed a $\chi^2$ minimization procedure, as described in detail for Procyon
in B15. For the \HST\/ data, we used the WFPC2 and WFC3 measures\footnote{For
the WFC3 measures, we used the ``F953N average'' values.} in Table~2, with the
(small) corrections to J2000 applied. 

We divided the measurements into three groups, based on the methods of
observation: visual micrometer; photographic, CCD, and mid-infrared (MIR); and
\HST\null. Then we assigned different uncertainties to each of the three data
sets, determined through an iterative procedure based on a comparison with the
simultaneous orbit fit, as follows. (1)~Few of the micrometer observations
contained explicit uncertainty estimates, so we set uniform uncertainties by
forcing the reduced $\chi^2_\nu$ to equal unity for this set of measurements. 
On the assumption that observations taken with a larger telescope are typically
more precise, and that observing techniques improved over time, we then scaled
the uncertainties proportionally to telescope aperture size by computing a
two-variable linear fit, and again forced $\chi^2_\nu$ to unity. (2)~For the
photographic and ground-based CCD+MIR measurements, we applied uniform
uncertainties that are isotropic in two dimensions to force the reduced
$\chi^2_\nu$ to 1 for this set of measurements. (3)~We did not scale the \HST\/
uncertainties, but left them at the values given in Table~2.

Following these adjustments to the measurement uncertainties of the first two
groups of data, we re-computed the fit to the entire set of ground-based and
\HST\/ measurements.  We used a sigma-clipping algorithm to reject measurements
for which the separation and/or PA was discrepant by more than four times the
standard deviation of the residuals.  We then re-scaled the uncertainties
according to the methods described above, and repeated the sigma-clipping
algorithm until no further measurements were rejected.  The final parameters for
the visual orbit are listed in Table~4. Uncertainties were computed from the
diagonal elements of the covariance matrix. The electronic table and associated
text in Appendix~A indicate which observations ended up being rejected.

For the \HST\/ data we found a $\chi^2$ per degree of freedom of
$\chi^2_\nu=1.87$. We believe this excess over unity is plausibly due to minor
systematic effects not included in our error determinations, such as changes in
plate scale due to the occasional adjustments in telescope focus, telescope
``breathing'' due to thermal effects in orbit, or small errors in the telescope
roll angle; or conceivably they could have an astrophysical origin (discussed
below in \S7).

We also experimented with a fit in which we fixed the orbital period based on
the entire set of data covering over 150 years, but then determined the
remaining parameters using only the high-precision \HST\/ data. However, this
resulted in the parameter uncertainties actually increasing by factors of about
1.6 to 4, and a reduction in $a$ by only 0.8~mas, which is small compared to its
uncertainty. Therefore we consider the values in Table~4 to be the best
estimates.

In Figure~3 (top panel) we plot all of the measurements, color-coded according
to their source (visual micrometer, photography, CCD and MIR, and \HST)\null.
The solid-line ellipse shows our orbital fit. For clarity, in the bottom panel
of Figure~3 we omit the numerous visual measures. The top panel shows that the
visual measurements frequently had very large errors, often of an arcsecond or
even more. There is a tendency, especially at the smaller separations, for the
separation to be systematically overestimated by the visual observers. The
bottom panel shows that even the CCD and MIR measures had a bias toward
overestimation of the separation. (Note that the semimajor axis had also been
systematically overestimated for Procyon in the historical data, as shown by
Girard et al.\ 2000 and B15.)

The observational errors for the \HST\/ data are too small to be visible in
Figure~3, so in Figure~4 we zoom in on a plot of the \HST\/ measures only ({\it
red dots}), along with the calculated positions ({\it open blue circles}) based
on our orbital solution. But once again the deviations from the orbital fit to
the exquisite \HST\/ astrometry are too small to be seen. 

\subsection{Residuals from Orbital Fit}

In Figure~5 we plot the residuals of our \HST\/ measurements as a function of
time relative to the orbital fit, in the directions of right ascension (top
panel) and declination (bottom panel). In right ascension the fit agrees with
all of the measures within 1$\sigma$ error bars, except for a $\sim$2.5$\sigma$
residual for a single observation. There appears to be no jump resulting from
the change of cameras in 2009. There is a slightly larger scatter in the
declination residuals, with two noticeable departures at the beginning and end
of 2006 of about 2$\sigma$ and 3.7$\sigma$. Because the PAs on those dates
were around $100^\circ$, a residual in declination of these sizes could in
principle be due to an incorrect telescope roll angle, at a level of about
$0\fdg08$. Normally the \HST\/ roll angle is known to about $\pm$$0\fdg003$, but
in very exceptional cases\footnote{See \tt
http://www.stsci.edu/hst/observatory/faqs/orient.faq} the uncertainty can
approach $0\fdg1$.

\section{Determining the Dynamical Masses}

\subsection{Parallax and Semimajor Axis of Sirius A}

In addition to the period, $P$, and the semimajor axis, $a$, of the relative
orbit listed in Table~4, we need two further quantities in order to determine
dynamical masses for both stars: the absolute parallax of the system, $\pi$, and
the semimajor axis, $a_A$, of the absolute orbital motion of Sirius~A on the
sky.

For the parallax, we used two independent determinations: (1)~GG78 calculated
the parallax from measurements of over 300 photographic plates obtained at the
Yerkes and Allegheny Observatories between 1917 and 1977, combined this result
with four earlier published determinations, and corrected the result from
relative to absolute to yield a parallax of $0\farcs3777\pm0\farcs0031$; (2)~the
absolute parallax was measured by the \Hipp\/ mission to be
$0\farcs37921\pm0\farcs00158$ (van Leeuwen 2007). These results are in good
agreement, and we adopt their weighted mean, as given in the top part of
Table~5. 

GG78 also determined the semimajor axis of Sirius~A's orbital motion from the
Yerkes and Allegheny material, and their result is presented in the bottom
section of our Table~5. However, the calculation needs to be repeated using our
new determination of the elements of the relative orbit. Fortunately GG78
tabulated the individual measurements of the photocenter location (their
Table~6), allowing us to repeat the determination of $a_A$; we found that one
observation, at 1966.824, was very discrepant, possibly due to typographical
error, and it was deleted. Our new adopted value of $a_A$ is given in the final
entry in Table~5.

\subsection{Dynamical Masses}

The final column in Table~6 lists the dynamical masses that result from our
adopted parameters. For comparison, the second column gives the masses derived
by van den Bos (1960), and the third column lists the values from GG78. We used
the usual formulae for the total system mass, $M = M_A+M_B = a^3/(\pi^3 \,
P^2)$, and for the individual masses, $M_A = M\,(1-a_A/a)$ and $M_B =
M\,a_A/a\,$; in these equations the masses are in $M_\odot$, $a$ and $\pi$ in
arcseconds, and $P$ in years. Our total system mass, $3.081\,M_\odot$, is about
3.6\% lower than was derived by GG78, and our individual masses,
$2.063\,M_\odot$ and $1.018\,M_\odot$, are smaller by similar
factors\footnote{If we derive orbital elements using {\it only\/} the \HST\/
astrometry, without the historical data, we find essentially the same masses but
with uncertainties $\sim$2.7 times larger: $2.068\pm0.062$ and
$1.020\pm0.030\,M_\odot$.}. These differences are due almost entirely to a 2.5\%
reduction in our adopted $a^3$, and a 1.0\% reduction in $\pi^{-3}$.  Compared
to the earlier study by van den Bos, our total system mass is lower by 3.4\%,
even though his value of $a=7\farcs50$ is very close to our measurement. In this
case, the difference is due mostly to our 3.1\% reduction in $\pi^{-3}$ relative
to the value used by van den Bos.

In Table~7 we present the error budgets for our derived masses of Sirius~A and
B, based on the random uncertainties of each of the parameters.\footnote{A
potential source of systematic uncertainty is errors in the plate scales of the
\HST\/ cameras. As we discussed in B15, Gonzaga \& Biretta (2010) state a
fractional uncertainty of $\pm$0.0003 for the WFPC2 plate scale, and for the
WFC3 plate scale we derived a similar fractional uncertainty of $\pm$0.00018.
These imply a systematic uncertainty of about $0\farcs0013$ for the semimajor
axis, $a$. Table~7 shows that a systematic error of this magnitude contributes
negligibly to the random errors in the dynamical masses.} As the table shows,
the mass errors are entirely dominated by the uncertainty in the parallax. 
Unfortunately, this error is unlikely to be reduced in the near future, because
Sirius is far too bright for its parallax to be measured by the current {\it
Gaia\/} mission (D.~Pourbaix and S.~Jordan, private communications).

\section{Astrometric Limits on Third Bodies}

The possibility that a third body exists in the Sirius system has been raised
many times. Early suggestions are reviewed by van de Kamp (1971), Greenstein et
al.\ (1971), Lindenblad (1973), and Brosch (2008), among others; they were based
on claims of direct visual detections of a companion object, or astrometric
perturbations of the binary orbit. Benest \& Duvent (1995) cited several
studies, including their own, that indicated a possible astrometric perturbation
of the visual orbit with a period of about 6~years. The (semi-)amplitude of the
claimed departure from two-body motion was found to be about $0\farcs055$. All
modern direct searches for a resolved companion using high-contrast imaging have
failed to reveal one down to limits corresponding to masses of several times
that of Jupiter (e.g., Schroeder et al.\ 2000; Bonnet-Bidaud et al.\ 2000;
Thalmann et al.\ 2011; Vigan et al.\ 2015; Bowler 2016; and references therein).
We likewise saw no evidence for a third object in any of our \HST\/ frames,
although they were not optimized for such a search.

The six-year orbital perturbation reported by Benest \& Duvent (1995), which
would produce offsets of over 100~mas peak-to-peak, is of an amount readily
detectable in our \HST\/ astrometry. However, our plots of the residuals in
Figure~5 show that the claim is categorically ruled out. As discussed above
(\S5.3), we see no evidence for perturbations in right ascension in excess of
the measurement uncertainties. The declination residuals (Figure~5 bottom
panel) tend to be slightly larger, and there was one noticeably large
(3.7$\sigma$) offset during late 2006, although not accompanied by a large
residual in right ascension. We calculated periodograms for the residual data,
but found no significant evidence for periodicities. In summary, although there
was one possible statistical fluke in late 2006, in declination only, we see no
convincing evidence for periodic perturbations with semi-amplitudes of more than
$\sim$5~mas. 

The long-term orbital stability of planets around individual stars in a binary
system has been studied numerically by, among others, Holman \& Wiegert (1999).
Using the results in their Table~3, and the parameters of the present-day
binary, we find that the longest periods for stable planetary orbits in the
Sirius system are about 2.24~yr for a planet orbiting Sirius~A, and 1.79~yr for
one orbiting Sirius~B\null.

We calculated the semimajor axes of the astrometric perturbations of both stars
that would result from being orbited by substellar companions of masses ranging
from 5 to $35\,\Mjup$ (where $\Mjup$ is the mass of Jupiter,
$0.000955\,M_\odot$), and for orbital periods up to the stability limits given
above. The results are plotted in Figure~6. For a semi-amplitude limit of 5~mas,
Figure~6 indicates that a companion of Sirius~A of $\sim\!15\,\Mjup$ or less
could escape astrometric detection. At $\sim\!25\,\Mjup$, only an orbital period
longer than $\sim$1~yr would have led to detection in our data. Progressively
more massive objects orbiting Sirius~A would have been detected more easily,
except at the shortest orbital periods. 

Our limits are more useful for Sirius~B, for which a precision RV study would be
impractical. A $\sim\!20\,\Mjup$ companion with a period longer than
$\sim$0.6~yr is excluded, but a $\sim\!10\,\Mjup$ satellite could in principle
be present at any period up to the stability limit. 

In summary, our findings are consistent with the tighter limits set by the
direct-imaging studies cited above.

\section{Astrophysics of the White Dwarf Sirius B}

We now turn to brief discussions of the astrophysical implications of our
precise dynamical masses for both stellar components of the Sirius system. We
begin in this section with the WD Sirius~B, and then consider the primary star,
Sirius~A, in \S9.

As discussed in \S1, Sirius~B is the nearest and brightest WD\null. It is a
nearly ideal target for astrophysical investigations---apart from its proximity
to the overwhelmingly bright primary, Sirius~A, making ground-based study of the
WD difficult. However, a wide range of high-quality space-based data has been
accumulated. These were summarized by Barstow et al.\ (2005), whose work has
been updated, using new observations, in a recent conference presentation
(Barstow et al.\ 2017). These studies are based in part on spectroscopic
observations of Sirius~B obtained with the Space Telescope Imaging Spectrograph
(STIS) on \HST\null. Unlike ground-based data, the STIS spectra have very minor
amounts of contamination by light from Sirius~A, easily subtracted to reveal a
high-SNR spectrum of Sirius~B\null. The spectra show only the Balmer lines,
confirming that the star is a DA2 WD with a pure hydrogen atmosphere.

Full details will be presented in a forthcoming journal paper, but our
model-atmosphere fitting to the STIS spectra yields an effective temperature and
surface gravity of $\Teff=25,369 \pm 46$~K and $\log g=8.591 \pm 0.016$ (cgs
units). Based on the absolute flux, the implied radius of Sirius~B is
$0.008098\pm0.000046\,R_\odot$ and its luminosity is
$0.02448\pm0.00033\,L_\odot$. (The quoted uncertainties are internal errors of
the model fits and do not include the modestly larger systematic uncertainties.)

In the two panels of Figure~7, we compare our measured parameters for Sirius~B
with theoretical predictions. We use theoretical modeling data from the
``Montreal'' photometric tables\footnote{{\tt
http://www.astro.umontreal.ca/$^\sim$bergeron/CoolingModels}. These tables are
based on evolutionary sequences and model atmospheres calculated by Holberg \&
Bergeron (2006), Kowalski \& Saumon (2006), Tremblay et al.\ (2011), and 
Bergeron et al.\ (2011).} for WDs with carbon-oxygen cores, pure-hydrogen
atmospheres, and a ``thick'' hydrogen layer with a mass of $M_{\rm
H}/M_{\Large*}=10^{-4}$. The top panel in Figure~7 shows the location of
Sirius~B in the theoretical Hertzsprung-Russell diagram (HRD; luminosity vs.\
effective temperature); the formal errors are smaller than the plotting symbol.
Also shown are the model cooling tracks for DA WDs with masses of 0.8, 0.9, 1.0,
and $1.2\,M_\odot$\null. If we were to infer the star's mass from its HRD
location---which was determined without reference to the mass---it would be
$1.019\,M_\odot$\null. This is in superb agreement with our measured dynamical
mass of $1.018 \pm 0.011 \, M_\odot$\null.  The top panel of Figure~7 also shows
isochrones for ages of 100, 150, and 200~Myr, again based on the Montreal
tables. By interpolation in the theoretical data, we estimate the cooling age of
Sirius~B to be 126~Myr. This agrees very well with an earlier determination of
124~Myr in a study of the Sirius system by Liebert et al.\ (2005, hereafter
L05).

In the bottom panel of Figure~7, we plot the position of Sirius~B in the
mass-radius plane, using our measured dynamical mass and the radius described
above. It is compared with a theoretical mass-radius relation for H-atmosphere
CO-core WDs with $\Teff = 25,369$~K, obtained through interpolation in the
Montreal tables. The observed mass and radius are in excellent agreement with
the theoretical relation. We show the dependence of the theoretical relations on
effective temperature by also plotting the curves for $\Teff = 10,000$ and
40,000~K; they clearly disagree with the observed values. To illustrate the
dependence on mean molecular weight, we additionally plot the Hamada \& Salpeter
(1961) mass-radius relation for zero-temperature WDs composed of $^{56}$Fe; it
is extremely discrepant with the observations.

The surface gravity of Sirius~B, calculated from our radius and dynamical mass,
is $\log g = 8.629 \pm 0.007$. This is only modestly discrepant (about
2.4$\sigma$) with the value of $\log g=8.591 \pm 0.016$ derived from the
model-atmosphere analysis of Barstow et al.\ (2017). The predicted gravitational
redshift (GR) is $79.8 \pm 1.0\,\kms$. This agrees very well with a measured GR
of $80.42 \pm 4.83 \,\kms$ reported in our earlier investigation (Barstow et
al.\ 2005). Unfortunately, in our more recent study of our new set of STIS
spectrograms we measure $89.60 \pm 0.75 \,\kms$. We are still investigating this
discrepancy and will discuss it in a separate forthcoming paper.

Based on an assumed initial-to-final-mass relation (IFMR), we can estimate the
initial mass of the Sirius~B progenitor, neglecting any interactions with
Sirius~A during its evolution. Adding the pre-WD evolutionary time of the
progenitor to the cooling time then yields the star's total age. We consider a
recent study by Cummings et al.\ (2016, hereafter C16), which presents two
versions of the IFMR\null. In the left panel of C16's Figure~8, they show an
IFMR  based on cluster ages calibrated using the Yonsei-Yale (Y$^2$) isochrones
database (Yi et al.\ 2001; Demarque et al.\ 2004), which only contains tracks
up to $5.0\, M_\odot$. In the right panel of the same figure, C16 show the IFMR
based instead on ``PARSEC'' (Bressan et al.\ 2012) isochrone data, which extend
to higher progenitor masses. The PARSEC calibration, although close to that of
Y$^2$ in the lower-mass region, changes slope for progenitor masses larger than
$4\, M_\odot$\null.  The uncertain location and shape of this break in slope
results in a progenitor-mass uncertainty for WDs in the vicinity of Sirius~B,
and thus an additional uncertainty in calculating the progenitor lifetime.

Assuming a slightly sub-solar metallicity for the progenitor of Sirius~B (see
next section), we performed two sets of evolutionary calculations, using the
Tycho and YREC codes (also described in the next section). When the PARSEC IFMR
is chosen, the estimated mass of the progenitor is
$5.6\pm0.6\,M_\odot$.\footnote{We corrected C16's PARSEC-based IFMR formula  for
a typographical error in the slope coefficient, from 0.097 to 0.0907 (J.
Cummings, private communication).} Using the Tycho evolution code, the
lifetimes from the zero-age main sequence (ZAMS) to the asymptotic giant branch
(AGB), are 132, 100, and 79 Myr, for progenitors of 5.0, 5.6 and $6.2\, M_\odot$,
respectively. These imply a total age of Sirius~B of $226^{+32}_{-21}$~Myr.

The Y$^2$-based IFMR calibration yields lower progenitor masses for Sirius~B,
and less observational scatter, in the range $5.0\pm0.2\, M_\odot$\null. In
order to reduce the systematic uncertainty inherent in combining data derived
with different evolution codes, we used an updated version of YREC to calculate
the progenitor lifetimes.  The derived lifetimes from the ZAMS were 112, 102,
and 94~Myr, for 4.8, 5.0, and  $5.2\, M_\odot$, respectively\footnote{The YREC
and Tycho codes produce different evolutionary timescales at masses around
$5\,M_\odot$, as discussed in more detail below in \S10.3.}. When added to the
cooling timescale estimate of 126 Myr, these lifetimes yield a total age for
Sirius~B in the range $228^{+10}_{-8}$~Myr.

\section{Astrophysics of Sirius A}

Our precise dynamical mass for Sirius~A ($2.063\pm0.023\,M_\odot$), and an age
estimate for the binary system ($\sim$226--228~Myr, with an uncertainty of about
$\pm$10~Myr) based on properties of the WD and an assumed IFMR, present an
opportunity to test theoretical models of the primary star's evolution.
Additional constraints are provided by determinations of the radius, luminosity,
and effective temperature of Sirius~A; for these we adopt the parameters given
by Davis et al.\ (2011, hereafter D11), as modified slightly by our adopted
parallax of $0\farcs3789$ instead of the $0\farcs37922$ used by D11. These
adjusted values are $R=1.7144\pm0.0090\,R_\odot$, $L=24.74\pm0.70\,L_\odot$, and
$\Teff=9845\pm64$~K\null. The radius of Sirius~A is tightly constrained, as it
is derived from precise interferometric measurements (e.g., Kervella et al.\
2003; D11; Boyajian et al.\ 2013 and references therein). A caveat, however, is
that these recent compilations (see also David \& Hillenbrand 2015 and Bohlin et
al.\ 2017) have given values of $L$ and $\Teff$ that range over several percent
relative to our adopted values. In order to test the agreement of models with
the observationally determined parameters for Sirius~A, we will compare with two
sets of theoretical evolutionary tracks calculated using the Yale Rotating
Stellar Evolution Code (YREC) and the Tycho\footnote{In spite of the YREC
acronym, all of the models in the work discussed here are non-rotating.
``Tycho'' is a name, in honor of Tycho Brahe, not an acronym.} code. 

YREC is a modern one-dimensional (1D) stellar-evolution code, designed to study
the hydrostatic phases of stellar evolution (Demarque et al.\ 2008). Convection
is included, using a solar calibration of mixing-length theory (MLT)\null. The
code has been continually updated; recent applications of YREC are given in
Spada et al.\ (2013) and Guenther et al.\ (2014). 

Tycho is also a 1D stellar evolution code, but it incorporates a description of
turbulent convection based on three-dimensional simulations of a convective zone
sandwiched between stable layers  (Arnett et al.\ 2015; Cristini et al.\ 2016).
These were analyzed using a Reynolds decomposition (Viallet et al.\ 2013), which
allows a quantitative and local determination of resolution errors. The errors
were small for the highest resolutions ($1536 \times 1024^2, 512^3, 1024^3$).
The description of convection has no adjustable free parameters. The simulations
exhibit narrow boundary layers and time-dependent turbulent entrainment,
features not found in MLT.

Differences in the location of Sirius~A in the HRD between models produced by
YREC and Tycho should be minimal. The effects of convection on the envelope
structure of an A-type star, which is primarily radiative, are small. Both YREC
and Tycho predict the presence of a convective core. The most significant
differences in the internal structure are the mass of the mixed core, and the
existence of a transition layer separating the convective core from the
radiative envelope, resulting in turbulent entrainment beyond the convective
boundary predicted by MLT. 

However, the codes also differ in the choice of the solar chemical composition,
which, through the requirement of consistency with the present-day Sun, affects
the opacities and equation of state. The YREC models use the solar mixture of
Grevesse \& Sauval (1998, hereafter GS98), chosen for its compatibility with
helioseismology (Basu \& Antia 2008). The addition of a boundary layer in the
Tycho models gives additional mixing, so that the solar model must compensate by
having lower opacity (lower metallicity). The Tycho models use the Lodders
(2010, hereafter L10) abundance tables, which have a lower metallicity than
GS98. The fractional abundances by mass of elements heavier than helium are
$Z=0.0169$ and 0.0141 in the GS98 and L10 mixtures, respectively.  Solar models
with self-consistent boundary layers are not yet available, but
asteroseismological data show encouraging agreement for the boundary layer in
more massive stars (e.g., $3.25\,M_\odot$; Arnett \& Moravveji 2017).

Using the YREC code, we ran a series of models of Sirius~A with the mass fixed
at the measured $2.06\,M_{\odot}$ and a range of metallicities. (Here, and in
the discussion below, we consider the bulk composition of the star, and neglect
any effect of the superficial metallic-line photospheric composition.) We were
unable to fit the location of Sirius~A in either the HRD or in a plot of $\log
L$ vs.\ $\log R$, under the assumption of solar metallicity, because Sirius~A is
too luminous. Instead, we were forced to use a modestly metal-poor composition;
these tracks have higher luminosities due to the reduced opacity of
metal-deficient stellar material. The three YREC tracks plotted using red lines
in the two panels of Figure~8 have metal contents bracketing the value giving
the best agreement, which has $Z=0.0124$ (corresponding to a logarithmic metal
deficiency of $\rm[Fe/H]=-0.13$ relative to the GS98 solar value).
Ages\footnote{As in \S8, ``age'' is calculated relative to the arrival of the
model on the ZAMS, and does not include the time spent in pre-main-sequence
evolution.} are marked with large dots along the tracks in steps of 100~Myr. 
The inferred age of Sirius~A, based on the $\log L$ vs.\ $\log R$ diagram, is
constrained to $237\pm15$~Myr. This is compatible, within the uncertainties,
with the total age of the WD Sirius~B that we obtained in \S8.

We also made Tycho runs for three $2.06\,M_{\odot}$ models with a range of metal
contents. Again, we find that the metallicity has to be slightly sub-solar in
order to reproduce the positions of Sirius~A in the HRD and $\log L$ vs.\ $\log
R$ diagrams. The green lines in the two panels of Figure~8 show the Tycho tracks
with $Z=0.0113$, 0.0120, and 0.0127, with the middle one giving the best fit
(corresponding to $\rm[Fe/H]=-0.07$ on the L10 scale). Note that the Tycho
models  solve for the compositional mixing separately from the structure
(``operator splitting''), which at present causes the small ``wiggles'' seen in
the plotted tracks. This allows efficient use of very large nuclear-reaction
networks.

The differences in slopes between the YREC and Tycho tracks seen in Figure~8
arise from the behavior of the convective core. The 321D algorithm implemented
in Tycho, which is based upon 3D simulations of turbulent convection and has no
parameters which must be fitted, predicts mixing beyond the convective
boundaries prescribed by MLT\null. The core thus remains larger and grows more
rapidly in the Tycho models, resulting in a more rapid increase in luminosity
with age. Ages are marked on the Tycho tracks, again in steps of 100~Myr. The
inferred age of Sirius~A is $247\pm15$~Myr.

Thus both the YREC and Tycho models imply that the bulk composition of Sirius~A
is slightly deficient in heavy elements with respect to the primordial
composition of the Sun, with effective values of [Fe/H] of about $-0.13$ to
$-0.07$. The absolute metallicity values agree extremely well between the
two codes, being $Z=0.0124$ and 0.0120, respectively. Because the photospheric
composition has been modified by diffusive and levitative processes, it does not
provide an observational test of the bulk abundances. As we noted earlier, many
heavy elements actually appear to be overabundant at the surface relative to the
Sun; for example Lemke (1989) found that iron itself has a photospheric
abundance of $\rm[Fe/H]=+0.25$. Both sets of tracks imply an age of Sirius~A of
about 237 to 247~Myr, with an uncertainty of approximately $\pm$15~Myr. These
results are in good agreement with the age of the binary inferred completely
independently in our discussion of the WD Sirius~B in \S8.

\section{Astrophysical Puzzles of the Sirius Binary System}

Sirius, the brightest star in the sky, and one of the nearest---far from being
well-understood---presents several astrophysical problems. These questions have
been discussed by L05, Brosch (2008), Bona{\v c}i{\'c} Marinovi{\'c} et al.\
(2008, hereafter BM08), Landstreet (2011, hereafter L11), Perets \& Kratter
(2012, hereafter PK12), and many others. In this section we review a few of
these issues in the light of our new findings.

\subsection{Have the Stars Interacted?}

In \S\S8--9, we discussed the two components of Sirius as if they have evolved
independently. But---even though we find consistent ages for Sirius~A and B
under this assumption---independent single-star evolution may not have been the
case. In the present-day orbit, the separation of Sirius A and B ranges from
31.5~AU at maximum to 8.1~AU at periastron. If the progenitor of B had a mass of
$\sim$5.0--$5.6\,M_\odot$, as discussed in \S8, then the total mass of the
system was reduced from an original $\sim$7.1--$7.7\,M_\odot$ to its present
$3.08\,M_\odot$, due to mass lost from the progenitor of B\null. Under the
assumption that this mass loss was isotropic and on a timescale long compared to
the orbital period (cf.\ Burleigh et al.\ 2002, \S2), and ignoring any
interactions between the stars, this implies that the periastron separation was
only $\sim$1.5--1.6~AU in the progenitor binary. This is smaller than the radius
attained in the AGB phase of a $\sim$5.0--$5.6\,M_\odot$ star, and thus the two
must almost certainly have interacted in the past. Yet the binary did not enter
into a common-envelope event, and the orbit did not even tidally circularize as
might have been expected---it still has an eccentricity of 0.59. Moreover, as 
discussed in \S\S8--9, Sirius~B appears to be a normal WD, and Sirius~A shows no
apparent departure from single-star evolution nor any obvious signs of a past
interaction.

Several authors, including BM08, have pointed out that Sirius is by no means
unique: there are many detached binary systems in which one component is a WD
that clearly interacted with the primary star in the past, yet they still have
eccentric orbits. For instance, barium stars, in which processed material from
an AGB star (now faded to a WD) is present on the surface of a companion, are
often in eccentric orbits (Izzard et al.\ 2010, and references therein). Another
example of a wide evolved binary that avoided circularization is the Procyon
system, consisting of an F5 subgiant and a DQZ WD companion in a 40.8-yr orbit,
with an eccentricity of 0.40 (see B15). BM08 (see also references therein for
earlier theoretical considerations) have modeled binary systems in which the
orbit is initially significantly eccentric before the more-massive component
reaches the AGB\null. Mass loss from the AGB star is enhanced at each periastron
passage, producing a growth rate in eccentricity larger than the rate of tidal
circularization. In \S4.1 of BM08 they explore parameter space to see whether
the Sirius system itself can be reproduced through this eccentricity-pumping
mechanism, and they are able to do so. The amount of material accreted by
Sirius~A from the progenitor of B in the successful scenarios is about
0.05--$0.1\,M_\odot$. In a later paper, L11 raised the possibility that as much
as $\sim$$0.5\,M_\odot$ was accreted by Sirius~A\null. However, a possible
objection to significant mass transfer from the companion is that the rotational
velocity of Sirius~A is small compared to most A-type stars (\S1), indicating
that it avoided being spun up by accretion to a short rotation period.

Since the photospheric composition of A appears to have been modified by
diffusive processes, it is difficult to apply strong constraints on accretion 
scenarios based on chemical abundances at its surface. However, L11, and earlier
Richer et al.\ (2000), suggested that the abundance patterns, in particular
deficiencies of C and O, and near-solar N, are at least qualitatively consistent
with accretion of CNO-processed material having occurred at the time Sirius~B
was an AGB star. The enrichments of the heavy elements Sr, Y, Zr, and Ba found
by L11 (3 to 30 times solar), and of Cu and heavier elements found by Cowley et
al.\ (2016; generally 10 to 100 times solar), may also be indicative of
accretion of $\it s$-process elements.  In addition, any mass accretion
might have resulted in helium enrichment of the atmosphere,  which would
increase element abundance ratios relative to hydrogen.

We made an exploratory theoretical study, in which an additional Tycho model was
run to assess whether a past accretion event could have produced a detectable
effect on the evolution of Sirius~A\null. A star initially of mass
$1.96\,M_{\odot}$, with a metallicity of $0.9\,Z_{\odot}$, was evolved  for
100~Myr, the approximate time for a $5.6\,M_{\odot}$ companion star to evolve to
the thermally pulsing AGB phase. At that point, steady accretion at a rate of
$10^{-7}\, M_\odot\,\rm yr^{-1}$ was implemented for a duration of 1~Myr,
yielding a final mass for the star of $2.06\,M_{\odot}$\null.  After the
accretion was terminated, the model readjusted on a thermal timescale, and then
converged onto an evolutionary track with a normal shape and rate of evolution.
This track had a slightly lower $\Teff$ and larger radius at a given luminosity
than a non-accreting model starting at a mass of $2.06\,M_{\odot}$\null. This
particular scenario results in a poorer fit to the observed stellar parameters
of Sirius~A than the single-star models discussed in \S9; however, the
unexceptional shape and evolutionary timescale of the post-interaction track
makes it difficult to rule out a similar accretion history for Sirius~A.

PK12 considered an alternative scenario, in which eccentric binaries with
compact components like Sirius are descended from systems that were initially
triple. The onset of mass loss as the most massive component evolves could
trigger an orbital instability, leading to ejections, interactions, or even
physical collisions. In \S7 of PK12, they consider the specific case of Sirius.
PK12 ran theoretical simulations of triple systems, with a third star orbiting
an inner binary having initially a small orbital eccentricity. They were able to
reproduce the current Sirius system in cases where the third star was eventually
ejected, and the inner binary pumped to high eccentricity. This scenario is, in
principle, testable, since it predicts existence of an ejected third star as a
very wide common-proper-motion companion.

\subsection{Does Sirius Belong to the Ursa Major Moving Group?}

Over a century ago, Hertzsprung (1909) pointed out that Sirius appears to share
the space motions of the bright A-type ``Dipper'' stars in Ursa Major. A
substantial literature has developed since then in which many authors have
assigned Sirius to membership in this Ursa Major Group (UMaG)\null. A detailed
investigation of the UMaG, and a summary of earlier work, is given in a
classical paper by Roman (1949). Eggen (1960) stated that Sirius is known to be
a member of the UMaG, indeed calling it the ``Sirius Group,'' or later (Eggen
1992) the ``Sirius Supercluster.'' Soderblom \& Mayor (1993) reviewed the
literature on moving groups up to 1993, and considered Sirius a ``probable''
member of the UMaG\null.

However, King et al.\ (2003, hereafter K03) made an updated examination of
stellar memberships in the UMaG, based on a large collection of new data,
including precise new parallaxes from the \Hipp\/ mission. This led to a
``clean'' sample of definite UMaG members, and demotion of Sirius to an
``uncertain'' membership category. In Figure~9 we plot the observational HRD
($M_V$ vs.\ $B-V$) for the ``certain'' UMaG members, using the data in K03
(their Table~5) for stars with a membership class of ``Y'' (i.e., ``yes,''
indicating definite membership). Also plotted are the data for Sirius~A, from
the same table. K03 derived an age of the UMaG of $500\pm100$~Myr based on Y$^2$
isochrones. We essentially verified this result  by using the Y$^2$ isochrones
database, together with the accompanying interpolation tool\footnote{Both
available for download from the Y$^2$ web page: \tt
http://www.astro.yale.edu\slash demarque/yyiso.htm} (Demarque et al.\ 2004), to
find a solar-composition isochrone that provides a reasonable fit to the UMaG
data; this is the 550~Myr isochrone plotted in Figure~9. The figure shows that
Sirius~A lies well to the left and below the main-sequence turnoff of the UMaG,
in accordance with the younger age implied by our discussion in \S\S8--9. We
calculated another Y$^2$ isochrone, for an age of 220~Myr and metallicity
$\rm[Fe/H]=-0.07$, consistent with our findings in \S9. This isochrone passes
directly through the Sirius point, as shown in Figure~9. We conclude that either
Sirius is too young to be a member of the UMaG, or that if it is a member, it is
conceivably a blue straggler or other exotic object. 

K03 quote a metallicity of $\rm[Fe/H]=-0.09$ for the UMaG, and in a more recent
paper Ammler-von Eiff \& Guenther (2009) state that UMaG members have [Fe/H]
lying in the range $-0.14$ to +0.06. The slightly sub-solar bulk metallicity of
Sirius~A that we deduced in \S9 is reasonably consistent with these values, so
it is difficult to exclude Sirius from the moving group solely on the basis of
its composition. 


\subsection{Sirius B and the Initial-Final Mass Relation}

In \S8 we adopted two versions of the IFMR from a recent paper by C16, in order
to obtain the initial mass of the progenitor of Sirius~B, from which we
determined its evolutionary timescale. Here we consider the reverse approach, in
which we constrain the total age of Sirius~B to be equal to that found in \S9
for the age of Sirius~A\null. Then we take the difference between that age and
the cooling age of the WD to be equal to the pre-WD evolutionary timescale for
the progenitor. Using our evolutionary codes, we finally infer the progenitor's
mass that yields this timescale.

The ages of Sirius~A from the YREC and Tycho codes are 237 and 247~Myr,
respectively (\S9). Taking the YREC age, and subtracting the 126~Myr cooling age
of Sirius~B (\S8), we find a pre-WD evolutionary timescale for the progenitor of
about 111~Myr. The YREC code gives this timescale for a progenitor mass of
$\sim\!4.8\,M_\odot$\null.

We also ran Sirius~B progenitor models using the Tycho code for masses of 4.85,
5, 5.1, 5.6, and $6.0\,M_{\odot}$, at a metallicity of $0.9\,Z_{\odot}$. These
simulations were terminated during the thermally pulsing AGB (TP-AGB), at which
point computational times become long. A conservative estimate is that
continuation of the evolution all the way through removal of the H envelope
would increase the measured ages by only $\sim$2\%. Including a 2\% adjustment
for the end of the TP-AGB, the lifetimes of these models, starting from the
ZAMS, are 145, 132, 112, 100, and 81 Myr, respectively. The age of Sirius~A,
based on the Tycho code, is 247~Myr. Subtracting the 126~Myr WD cooling age
yields a progenitor pre-WD timescale of 121~Myr, corresponding to a progenitor
mass of $5.06\,M_\odot$.  

There is a significant discrepancy in stellar lifetimes in this mass range
between Tycho and YREC\null. It arises from the treatment of convection. The
additional turbulent entrainment beyond the thermodynamic boundary of the
convective core that is a feature of the 321D algorithm in Tycho is predicted to
increase with stellar mass, resulting in longer main-sequence lifetimes and
larger He cores. The final mass of the WD in Tycho is also larger than in YREC
for a given progenitor mass, due to the larger He core and the larger extent of
convection during He burning. Thus the relatively small differences in the
$2.06\,M_{\odot}$ Sirius~A models become much more marked in a $5\,M_{\odot}$
model. 

We note that the exact final WD mass predicted by the Tycho code depends on
details of the TP-AGB evolution, but the models allow us to constrain the
progenitor mass needed to yield a $1.018\,M_{\odot}$ WD at this metallicity to
about $5.0\pm0.1\, M_{\odot}$, in agreement with the lifetime argument above.

Our result ($M_{\rm initial}=4.8$--$5.06\,M_\odot$ and $M_{\rm
final}=1.018\,M_\odot$) lies well within the observational scatter for WDs in
open clusters shown in the IFMR of Figure~8 in C16. Our results are also
consistent with the earlier study of L05, who inferred a Sirius~A age of
225--250~Myr, and an initial mass for Sirius~B of
$5.06^{+0.37}_{-0.28}\,M_\odot$. Thus, again, we see no direct evidence for a
departure from normal single-star evolution.

\section{Summary }

Based on our analysis of nearly two decades of precise astrometry of the Sirius
system with the {\it Hubble Space Telescope}, ground-based photographic
observations presented here for the first time, and historical measurements
dating back to the 19th century, we have derived dynamical masses for both
components. The metallic-line A star Sirius~A is  found to have a mass of $2.063
\pm 0.023 \, M_\odot$, and the Sirius~B WD companion has a mass of $1.018 \pm
0.011 \, M_\odot$. In spite of past claims, we find no evidence for
perturbations due to third bodies in the system, at levels down to masses of
about 15--$25\,M_{\rm Jup}$\null. 

The position of Sirius~B in the H-R diagram is in excellent agreement with a
theoretical cooling track for a WD of its measured mass, and implies a cooling
age of 126~Myr. In the mass-radius plane, Sirius~B's location is likewise in
agreement with theoretical predictions for a carbon-oxygen white dwarf of its
effective temperature, with a hydrogen-dominated atmosphere. 

We calculated evolutionary tracks for stars with the mass of Sirius~A, using two
modern codes. In order to fit the observed parameters (radius, luminosity, and
effective temperature) we find it necessary to adopt a slightly subsolar bulk
metallicity, of about $0.85\,Z_\odot$. The two codes yield ages for Sirius~A in
the range of about 237--247~Myr. This age range is consistent with the age of
Sirius~B, if we add a plausible pre-WD evolutionary timescale to its cooling
age.

In spite of the apparent consistencies with the assumption that the two stars
have evolved independently, we point out that the binary might have been closer
in the past, before the progenitor of Sirius~B underwent significant mass loss.
Thus it is difficult to understand how they could have avoided an interaction
and mass accretion onto Sirius~A\null. There are indeed tantalizing hints in the
photospheric composition of Sirius~A for contamination from an AGB wind or
Roche-lobe overflow, but the evidence is obscured by apparent levitative
processes in the stellar atmosphere. The slow rotational velocity of Sirius~A
and the high eccentricity of the present-day orbit are also problematic for a
scenario involving a past interaction. 

We considered the long-standing claim that Sirius belongs to the Ursa Major
moving group, with which it appears to share a common space motion. However,
Sirius~A has the appearance of being significantly younger than the group
members, perhaps indicating that it simply is not a physical member.
Alternatively, the seemingly well-behaved Sirius system may be concealing an
exotic past evolutionary history involving interactions and mass transfer
between the two stars, or even one necessitating a third star that was
dynamically ejected from the system while exciting the remaining binary to a
higher orbital eccentricity. The brightest star in the sky continues to stand as
a beacon challenging our understanding of stellar evolution.

\acknowledgments

We acknowledge the contributions of over 130 dedicated and patient observers of
the Sirius system since the discovery of Sirius~B in 1862. Support was provided
by NASA through grants from the Space Telescope Science Institute, which is
operated by the Association of Universities for Research in Astronomy, Inc.,
under NASA contract NAS5-26555. G.H.S. acknowledges support from National
Science Foundation grant AST-1411654, and J.B.H. from NSF grant AST-1413537. 
This research has made use of the Washington Double Star Catalog maintained at
the USNO. We thank Jeff Cummings for a useful discussion.

{\it Facilities:} \facility{\HST\/ (WFPC2, WFC3)}, \facility{USNO:26-inch}

\appendix

\section{Critical Compilation of Historical Observations of the Sirius Binary
System}

\begin{center}
{\it By Gail H. Schaefer, Jay B. Holberg, and Brian D. Mason}
\end{center}

We have assembled what we believe to be a complete compilation of all published
historical measurements of the position angle (PA) and angular separation of
Sirius B relative  to the primary star. Our tabulation is based on a critical
review of measures  contained in the Washington Double Star Catalog
(WDS)\footnote{{\tt http://ad.usno.navy.mil/wds}. Sirius is designated WDS
J06451$-$1643 in the WDS catalog.} maintained at the USNO, and from our
additional literature searches. 

The complete tabulation is presented in the electronic version of this paper,
and will also be available at the VizieR website.\footnote{\tt
http://vizier.u-strasbg.fr/viz-bin/VizieR} Notes at the end of the tabulation
give extensive commentary on the historical observations.  Table~8 in the
present paper shows an excerpt from the full table, with some of the columns
omitted for clarity. The table presents the date of observation (Besselian
year), measured PA and separation, PA and separation corrected to J2000 (as
described in \S5.1) and their adopted uncertainties, an observer code, a code
for method of observation, telescope aperture, and a code for notes and remarks.
A second tabulation contains the bibliographic codes (BibCodes) and full
literature references corresponding to the observer codes; an excerpt from this
tabulation is given in Table~9.

The visual micrometer observations did not always include a contemporaneous
measurement of both PA and separation. In Table~8, these omissions are listed
with a value of $-99.0$ and the associated errors are set to zero. The adopted
uncertainties were determined as described in \S5.2, but are listed with a value
of 0.0 for measurements that were rejected from the orbit fit.

Many early publications provided measures averaged over multiple nights or even
an entire observing season, for the purpose of reducing computational labor in
subsequent analyses.  With modern computers there is no need for such averaging,
so we opted to present the individual measures whenever available. However, if
an observer reported more than one measurement on a given night, we did compute
the mean position for that night. If the original publication only reported a
mean across several nights, we tabulate that mean as reported.  

Some early results were reported in more than one publication.  We identified
these by listing additional reference codes in the online version of the table. 
There were several cases where the results were slightly different from one
publication to another; we identify these instances in the notes column, and
provide an explanation for selecting the values listed in the table.  Here we
discuss one specific case in more detail: Struve (1893) attempted to correct his
measurements for systematic errors by measuring artificial double stars, and he
published both the original and corrected measures for the companion of Sirius.
Aitken (1935, p.~61) cautions that the variances of the PAs and separations
compared with those of other observers are often larger for Struve's corrected
values than for the originals.  However, in the case of Struve's measurements of
Sirius, we in fact found that the mean residuals in the PAs did improve when we
used Struve's corrected values, while the separation residuals did not change
significantly; we therefore chose to use Struve's corrected values.

Another set of measurements that we discuss in more detail are the USNO
photographic observations by Lindenblad (1970, 1973).  He published data for the
raw individual measures, but only corrected the seasonal means for emulsion
contraction (see \S4.1). The average time span covered by his mean measures is
$\sim$70 nights. We chose to apply Lindenblad's tabulated scale factors in order
to compute corrected separations and PAs for the individual measurements.  We
then averaged the measures for plates taken on the same night, reducing the
number of individual measurements from 157 down to 77 observations on unique
dates.

The electronic table contains a total of 2354 measurements. Of these, two
observations by G.~M. Searle at 1866.93 and 1867.03 (1882, Harvard Annals, 13,
p.~36) were rejected because his notes indicate that they do not refer to
Sirius~B\null. Additionally, the two {\it HST\/} measurements by Schroeder et
al.\ (2000) were replaced by our reanalysis of the same images, as described in
\S3. The table also includes eight attempted observations wherein the binary was
unresolved, or where no formal measurement of the binary position was reported,
including the initial discovery by the Clarks.

A total of 2350 measurements remains for inclusion in our initial orbital fit,
contributed by 135 distinct observers. There are 1915 visual micrometer
observations, 407 photographic observations, and only 28 using ``modern''
techniques. Remarkably, aside from the \HST\/ observations reported here,
astrometry of Sirius~B has been almost entirely neglected by professional
astronomers for more than the past three decades, subsequent to a final 1986
photographic observation reported by Jasinta \& Hidayat (1999). The only
exception of which we are aware is a single measurement in 2005 derived from MIR
observations with the Gemini South telescope (Skemer \& Close 2011). Ten
measurements by two amateur astronomers based on CCD frames obtained with small
telescopes between 2008 and 2016 have been reported, their most recent
publications being Anton (2014) and Daley (2016).

In our orbital solution described in \S5.2, we used a sigma-clipping algorithm
to reject badly discrepant measurements. We rejected a total of 67 observations:
59 micrometer, five photographic, and three amateur CCD measurements.  The
rejected observations are identified in the Notes column of Table~8.  As
mentioned in the notes to the electronic version, many of these observations
were of dubious quality to begin with.


\clearpage


\begin{figure}
\begin{center}
\includegraphics[width=5in]{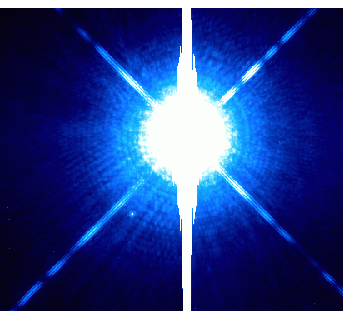}
\figcaption{False-color rendition of an \HST\/ image of Sirius, from a WFPC2
frame obtained in the near-infrared F1042M filter on 2001 October~27, exposure
time 35~s.  The white dwarf Sirius~B lies to the lower left of the grossly
overexposed Sirius~A, at a separation of $5\farcs191$. The diffraction spikes
were used to locate the centroid of Sirius~A, as described in the text. In this
near-IR bandpass the brightness difference is about 10.6~mag. 
}
\end{center}
\end{figure}

\begin{figure}
\begin{center}
\includegraphics[width=4in]{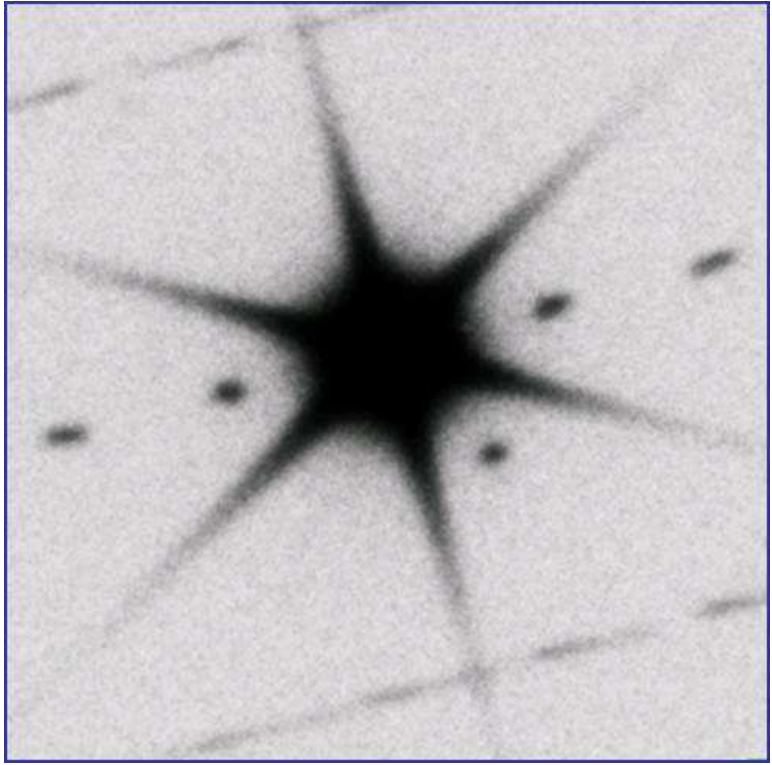}
\figcaption{
Digitized image of Sirius taken from a scan of a photographic plate obtained
with the USNO 26-inch refractor. A hexagonal mask was used in front of the
objective, and oriented to place Sirius~B between two of the spikes at the lower
right. An objective wire grating was also used, producing first- and
second-order images on either side of the overexposed Sirius~A, from which its
centroid can be determined.
}
\end{center}
\end{figure}

\begin{figure}
\begin{center}
\includegraphics[width=4.25in]{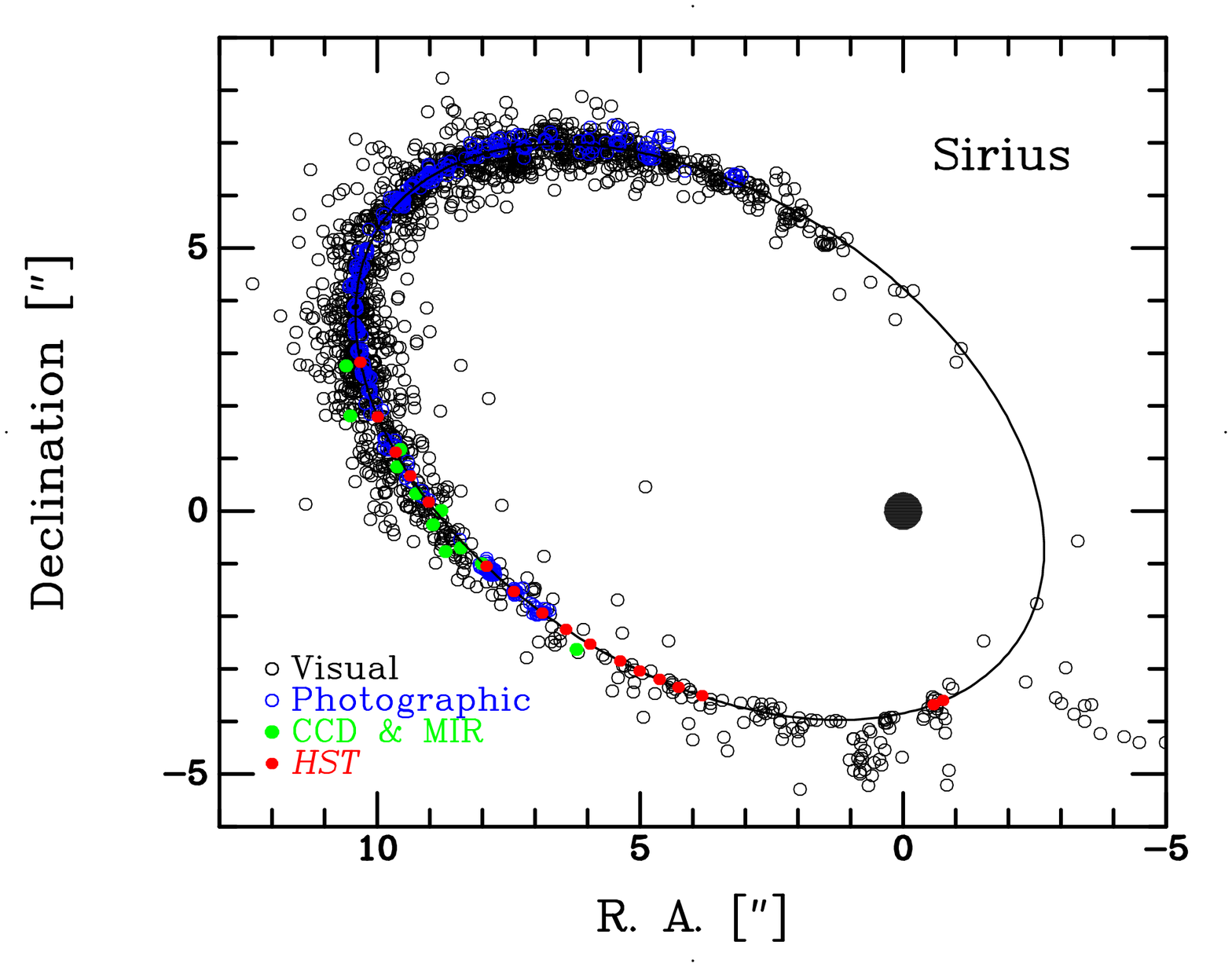}
\vskip 0.1in
\includegraphics[width=4.25in]{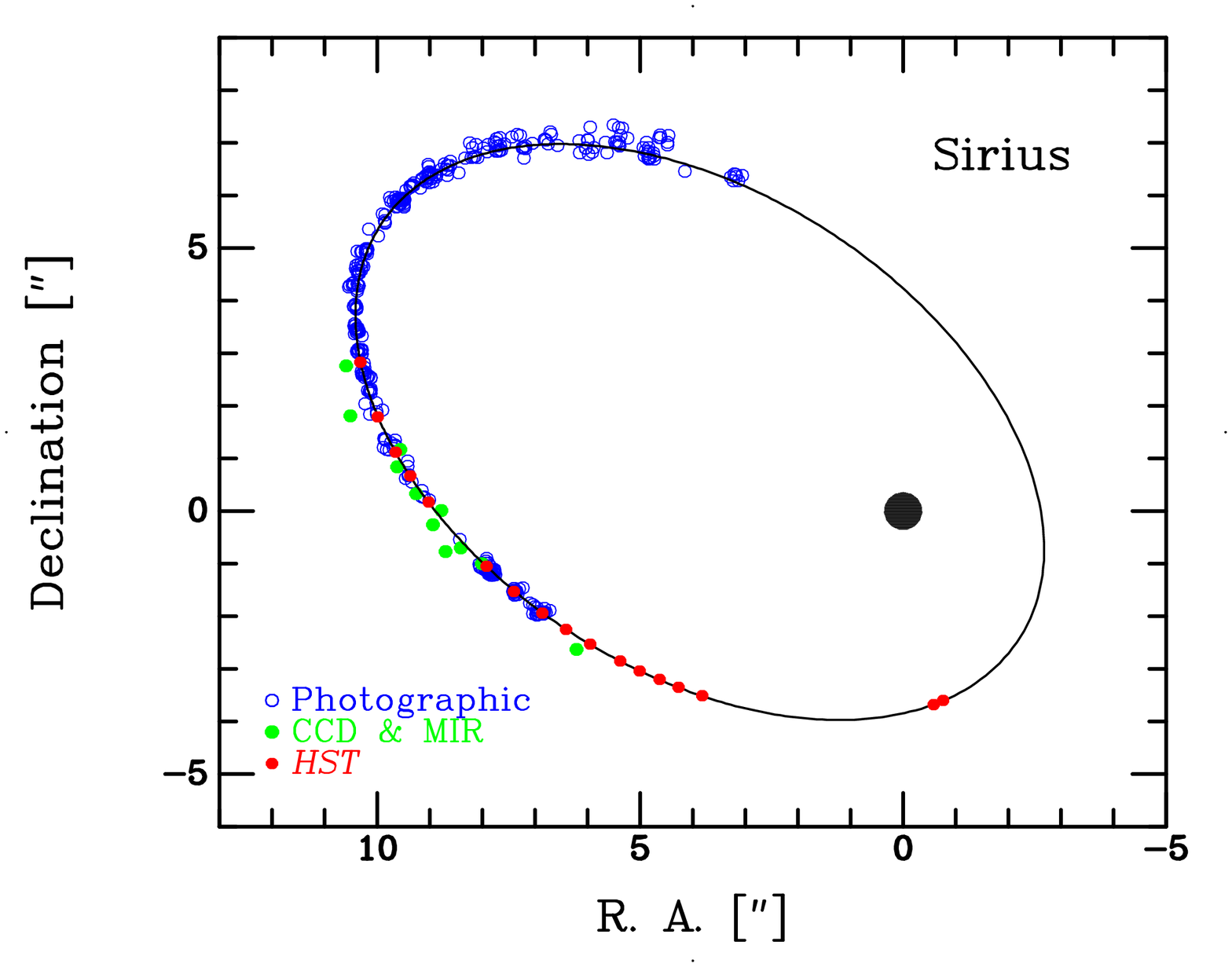}
\end{center}
\vskip-0.4in
\caption{
{\bf Top:} 
Historical and \HST\/ observations of the relative orbit of Sirius~B\null.
Visual micrometer measures are plotted as {\it open black circles},
photographic as {\it open blue circles}, CCD and mid-IR as {\it filled green
circles}, and \HST\/ as {\it filled red circles}. The {\it black ellipse\/}
plots our orbital fit from \S5.2. 
{\bf Bottom:}
Same figure, but with the visual micrometer observations omitted in order to
show the other measures more clearly. 
}
\end{figure}

\begin{figure}
\begin{center}
\includegraphics[width=6.5in]{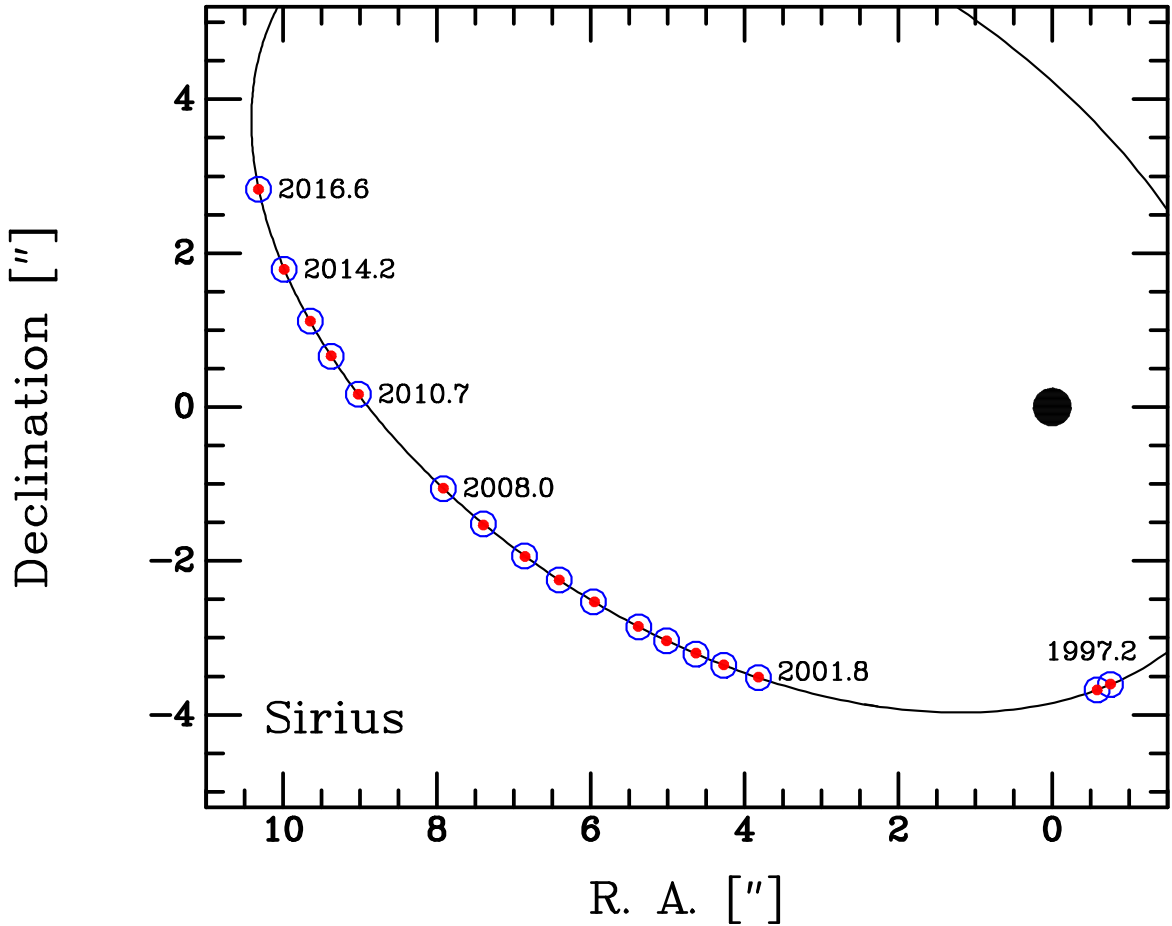}
\end{center}
\vskip-0.4in
\caption{
Close-up view of \HST\/ observations of the relative orbit of Sirius~B, plotted
as {\it filled red circles}. Dates of the observations are indicated for a few
of them. The {\it open blue circles\/} plot the calculated positions based on
our final orbital fit, whose elements are given in Table~4. The orbit is
plotted as the {\it black ellipse}.
}
\end{figure}

\begin{figure}
\begin{center}
\includegraphics[width=4.5in]{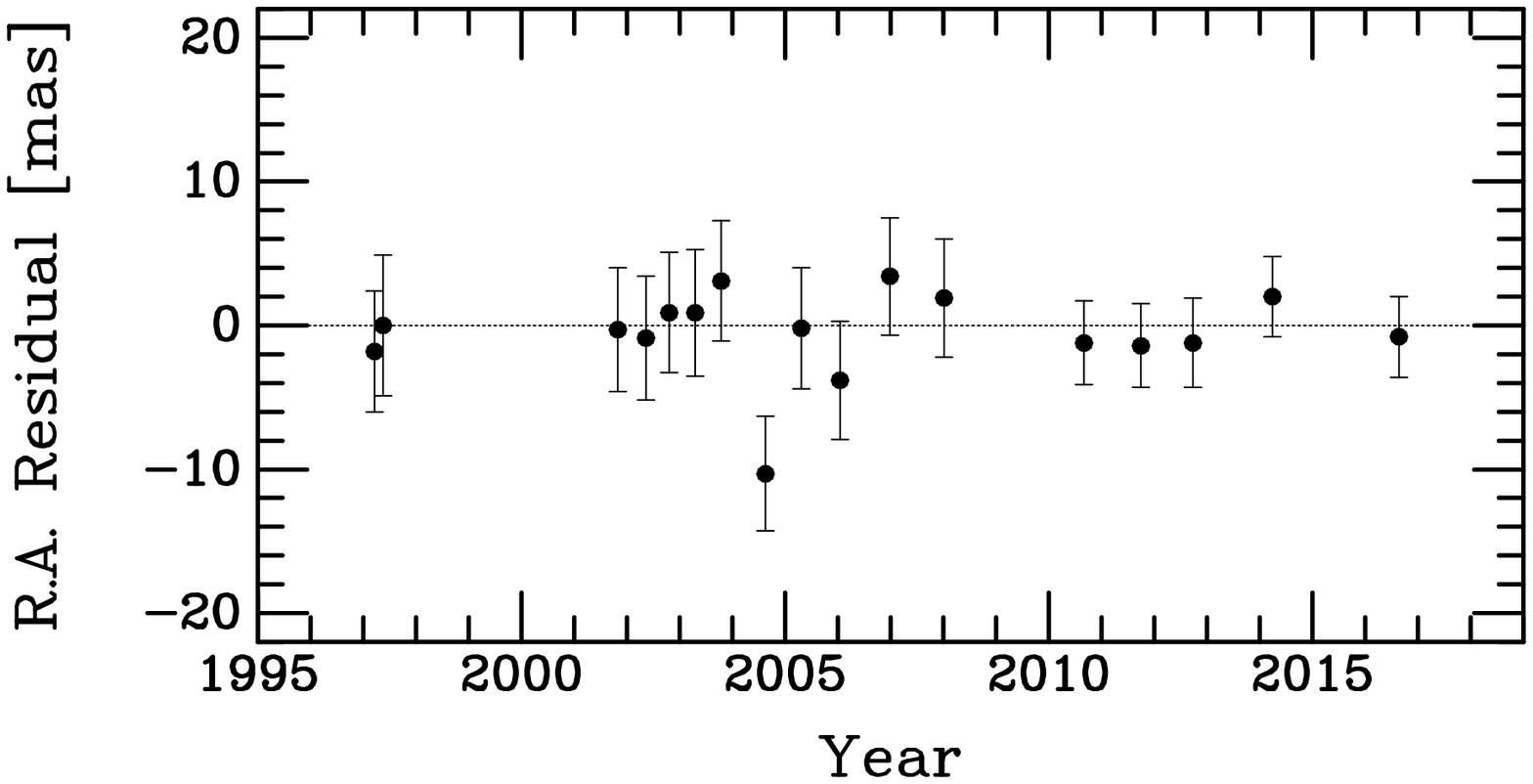}
\vskip 0.25in
\includegraphics[width=4.5in]{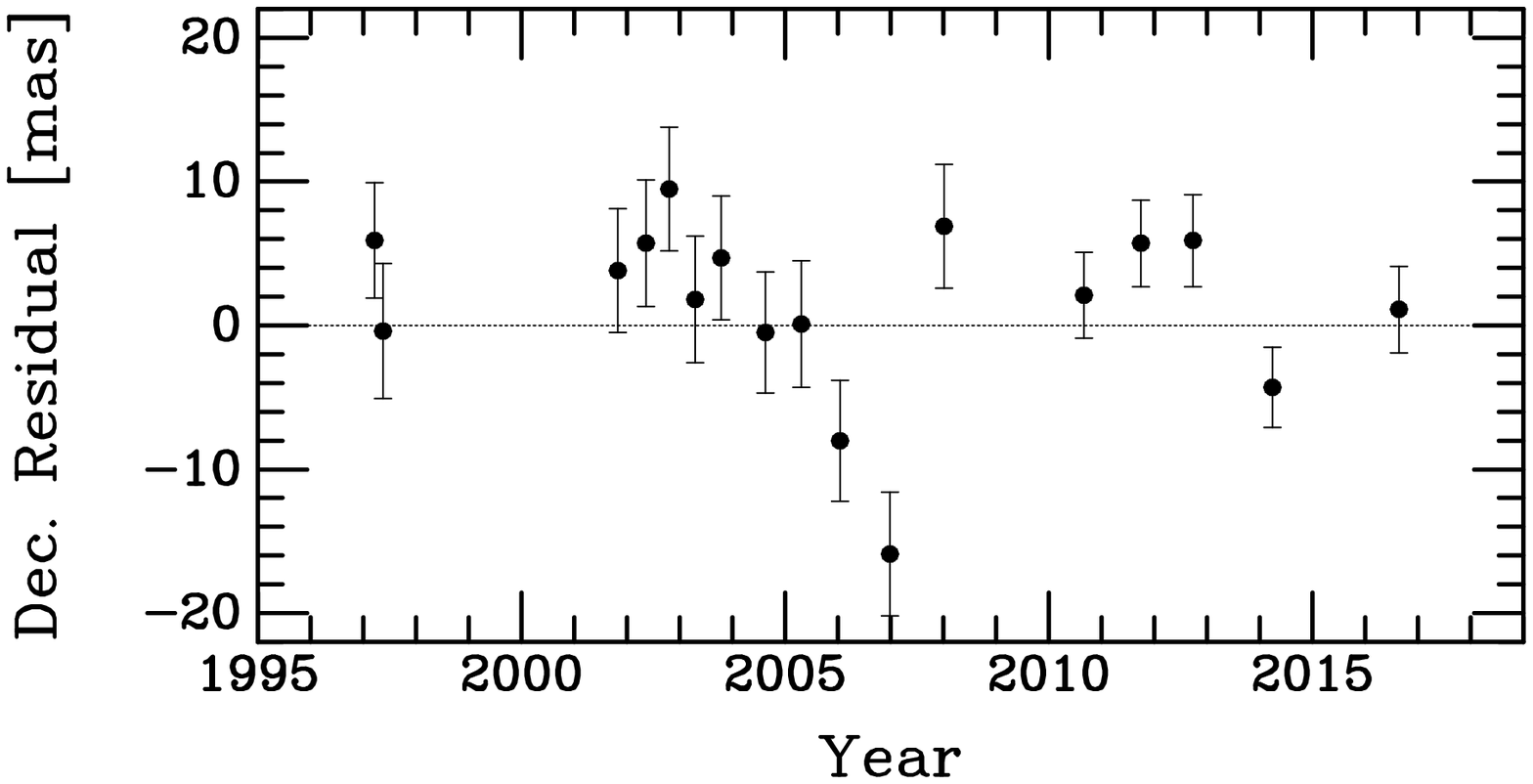}
\end{center}
\vskip-0.2in
\caption{
Residuals (in milliarcseconds) between the right-ascension ({\bf top panel}) and
declination ({\bf bottom panel}) position offsets of Sirius~B from Sirius~A
observed with \HST, and the offsets predicted by our adopted orbital elements.
}
\end{figure}

\begin{figure}
\begin{center}
\includegraphics[width=5.5in]{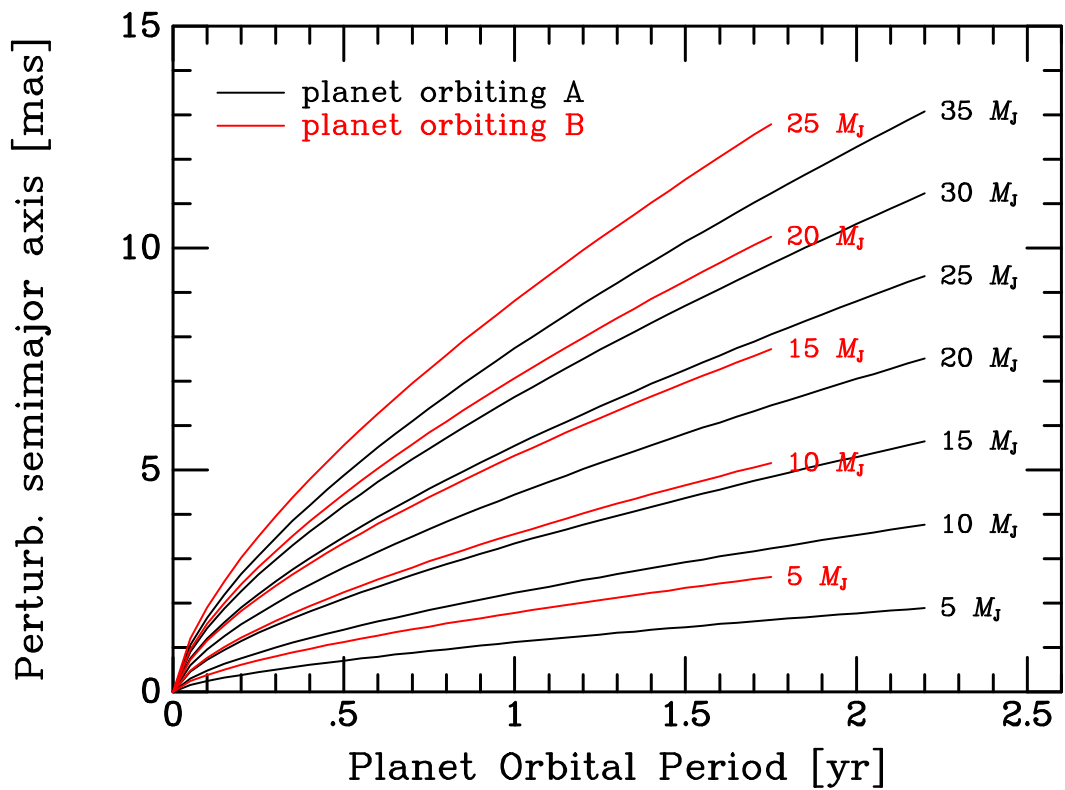}
\end{center}
\vskip-0.2in
\caption{
Astrometric perturbations that would result from planetary companions of
Sirius~A ({\it black curves}) or Sirius~B ({\it red curves}), with the masses
of the perturbers (in units of the Jovian mass) indicated in the labels.
Calculations were made for periods up to the orbital-stability limits of planets
with orbital periods of $\sim$2.24~yr (companions of Sirius~A) or $\sim$1.79~yr
(companions of Sirius~B). The $y$-axis is the semimajor axis of the resulting
astrometric perturbation of A or B in milliarcseconds.
}
\end{figure}

\begin{figure}
\begin{center}
\includegraphics[width=3.95in]{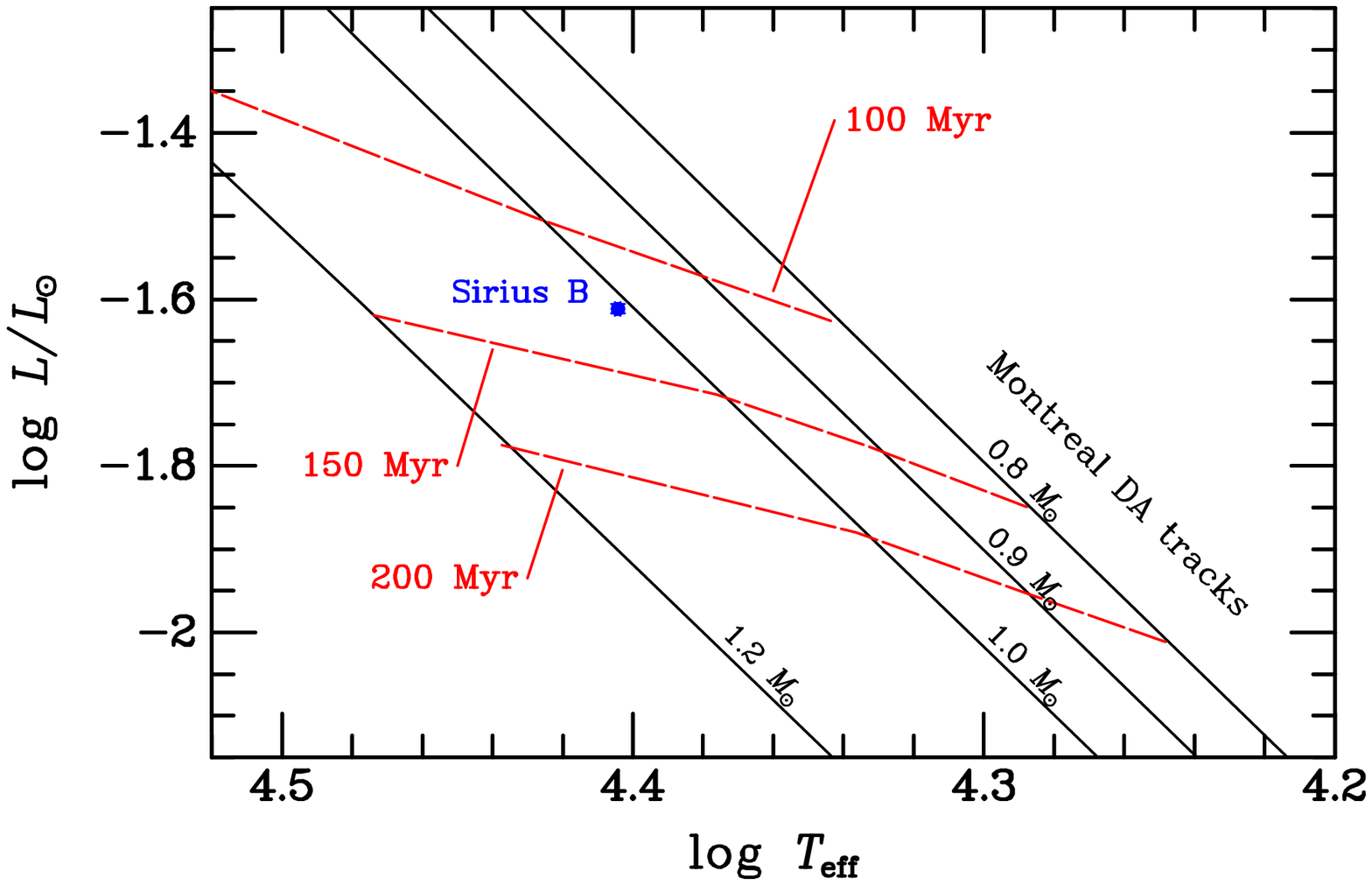}
\vskip 0.2in
\includegraphics[width=3.95in]{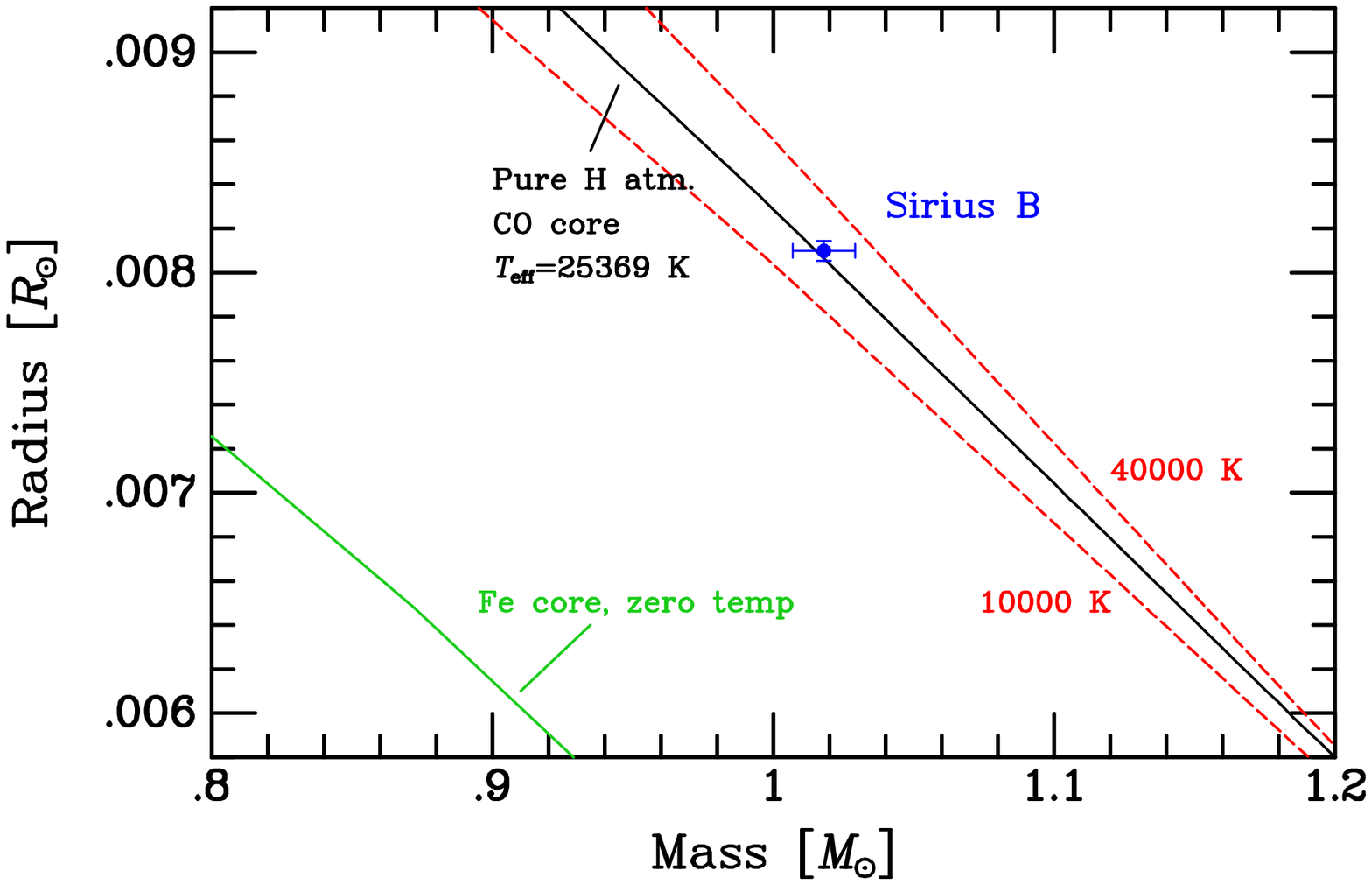}
\end{center}
\vskip-0.2in
\caption{
Comparisons of white-dwarf theory with the observed parameters of Sirius~B\null.
{\bf Top:} Observed position of Sirius~B in the theoretical H-R diagram,
compared with Montreal cooling tracks ({\it black lines}) and isochrones ({\it
dashed red lines}) for pure hydrogen-atmosphere CO-core white dwarfs of the
indicated masses. The implied mass is $1.019\,M_\odot$, in excellent agreement
with the measured $1.018\,M_\odot$. The inferred white-dwarf cooling age of
Sirius~B is 126~Myr. 
{\bf Bottom:} Observed position of Sirius~B in the mass-radius plane, compared
with a theoretical relation ({\it black line}) for pure H-atmosphere CO-core
white dwarfs of effective temperature $\Teff=25,369$~K, based on the Montreal
database. Shown as {\it dashed red lines\/} are the relations for CO white
dwarfs with $\Teff=10,000$ and 40,000~K\null. Also plotted ({\it green line}) is
the Hamada--Salpeter mass-radius relation for a zero-temperature white dwarf
composed of iron. The agreement of theory with observations is excellent,
verifying that Sirius~B is a CO-core white dwarf. 
}
\end{figure}

\begin{figure}
\begin{center}
\includegraphics[width=4.25in]{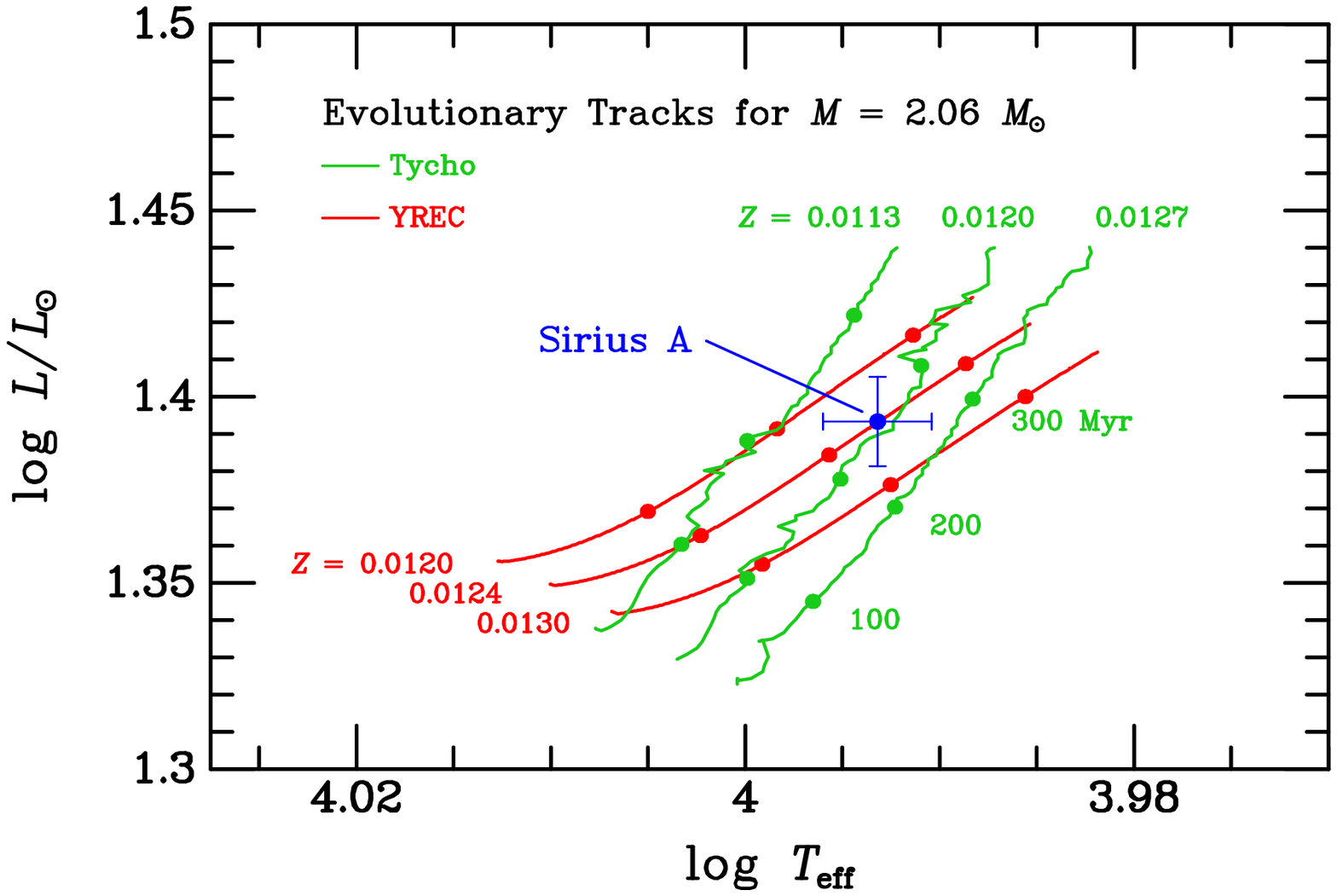}
\vskip 0.2in
\includegraphics[width=4.25in]{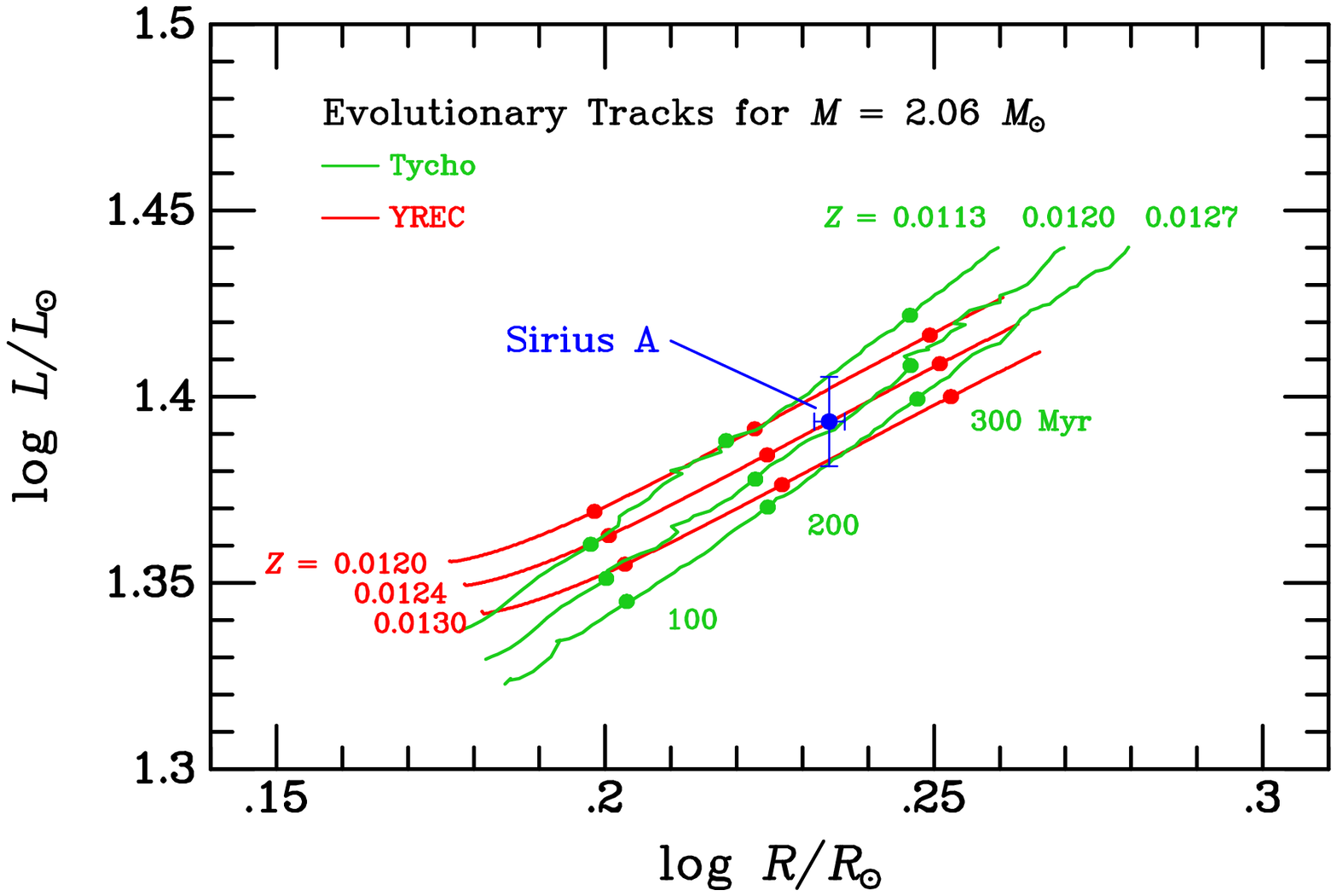}
\end{center}
\vskip-0.2in
\caption{
Theoretical evolutionary tracks in the H-R diagram {\bf (top panel)} and in
$\log L/L_\odot$ vs.\ $\log R/R_\odot$ {\bf (bottom panel)} for models with
masses set to that of Sirius~A, $2.06\,M_\odot$. The {\it green curves\/} plot
models calculated with the Tycho code, and the {\it red curves\/} plot YREC
tracks. Heavy-element contents by mass, $Z$, are indicated in the figures. The
observed position of Sirius~A is plotted as {\it blue points\/} with error bars.
In both evolutionary codes, the parameters of Sirius~A are reproduced with
slightly sub-solar metallicities of about $Z=0.0120$--0.0124. The {\it green and
red dots\/} on the tracks mark ages of 100, 200, and 300~Myr. Both codes
indicate an age of $\sim$237--247~Myr for Sirius~A.
}
\end{figure}

\begin{figure}
\begin{center}
\includegraphics[width=5.5in]{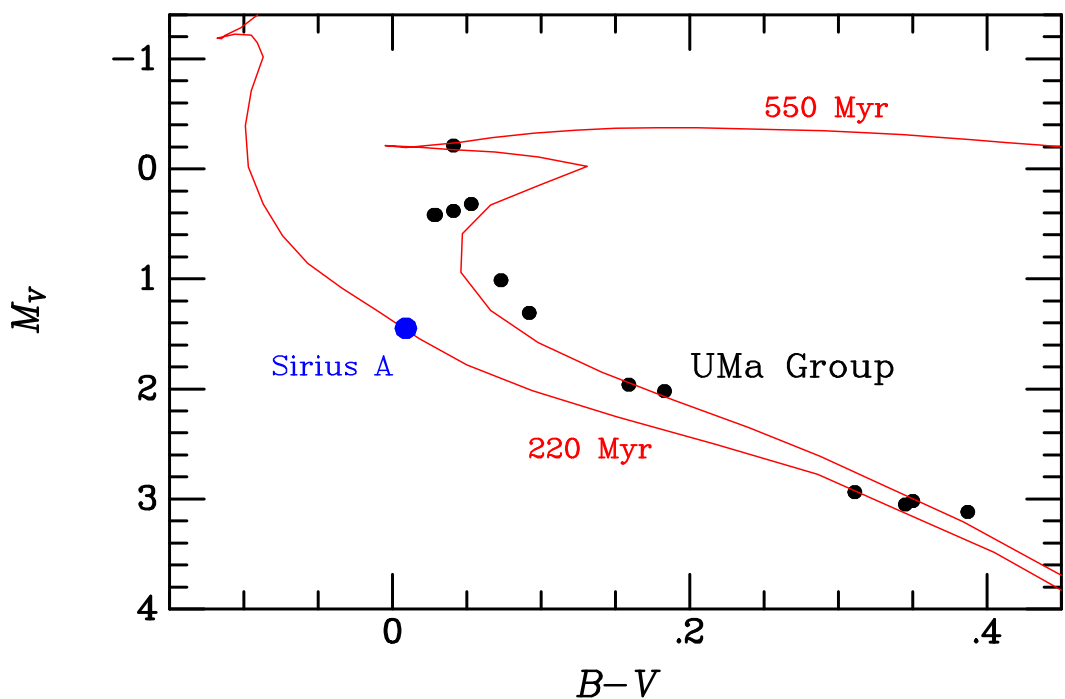}
\end{center}
\vskip-0.2in
\caption{
Color-magnitude diagram (absolute $V$ magnitude vs.\ $B-V$ color) for Ursa Major
group certain members ({\it filled black circles}) and Sirius~A ({\it filled
blue circle}); data taken from King et al.\ (2003). The two {\it red lines\/}
show isochrones for 550~Myr age with solar composition, and for 220~Myr age with
$\rm[Fe/H]=-0.07$; both were generated using the Y$^2$ isochrones interpolation
tool (see text). The location of Sirius~A to the left of the Ursa Major main
sequence turnoff suggests it to be younger than the group, thus calling into
question its group membership---unless it is a blue straggler mimicking a young
age.
} 

\end{figure}

\clearpage


\begin{deluxetable}{lllcc}
\tablewidth{0 pt}
\tabletypesize{\footnotesize}
\tablecaption{\HST\/ Observing Log for Sirius}
\tablehead{
\colhead{UT Date} &
\colhead{Dataset\tablenotemark{a}} &
\colhead{Exposure } &
\colhead{No.} &
\colhead{Proposal} \\
\colhead{} &
\colhead{} &
\colhead{Times [s]\tablenotemark{b}} &
\colhead{Frames\tablenotemark{c}} &
\colhead{ID} 
}
\startdata
\multispan5{\hfil WFPC2/PC Frames, F1042M Filter \hfil} \\
1997 Mar 19 & u3mi1503r & 12, 100    & 8  & 6887   \\
1997 May 18 & u3mi1603m & 12, 100    & 8  & 6887   \\
2001 Oct 27 & u6gb0202m & 4, 5, 6, 7, 35 & 10 & 9072   \\ 
2002 May 10 & u6gb0306m & 8, 40, 60  & 10 & 9072   \\ 
2002 Oct 20 & u8if0206m & 8, 40, 60  & 10 & 9334   \\ 
2003 Apr 18 & u8if0306m & 8, 40, 60  & 10 & 9334   \\ 
2003 Oct 15 & u8tp0206m & 8, 40, 60  & 10 & 9964   \\ 
2004 Aug 15 & u8tp0301m & 8, 60      & 12 & 9964   \\ 
2005 Apr 20 & u8tp0601m & 8, 60      & 12 & 9964   \\ 
2006 Jan 15 & u9bv0101m & 8, 60      & 12 & 10619  \\ 
2006 Dec 27 & u9o60101m & 8, 40, 60  & 13 & 10990  \\ 
2008 Jan 03 & u9z80101m & 8, 60      & 12 & 11290  \\ 
\multispan5{\hfil WFC3/UVIS Frames, F953N Filter \hfil} \\
2010 Sep 02 & ibk703010 & 6, 12      & 28 & 12296  \\ 
2011 Oct 01 & ibti03010 & 6, 12      & 28 & 12673  \\  
2012 Sep 26 & ic1k03010 & 6, 12      & 28 & 13062  \\  
2014 Mar 31 & ica103010 & 6, 12      & 28 & 13468  \\  
2016 Aug 20 & icvd03010 & 6, 12      & 28 & 14342  \\
\enddata
\tablenotetext{a}{Dataset identifier for first useful observation made at each
visit. Principal Investigator was H.~Ford for first two epochs, H.E.B. for the
rest.
}
\tablenotetext{b}{Exposures of 0.11~s were also taken during some WFPC2 visits,
but were not used in our astrometric analysis.
}
\tablenotetext{c}{Total number of useful individual frames obtained during each
visit.
}
\end{deluxetable}

\begin{deluxetable}{ccccl}
\tablewidth{0 pt}
\tabletypesize{\footnotesize}
\tablecaption{{\em HST\/} Astrometric Measurements of Sirius~B Relative to
Sirius~A}
\tablehead{
\colhead{UT Date} &
\colhead{Besselian} &
\colhead{Separation} &
\colhead{J2000 Position} &
\colhead{Source} \\
\colhead{} &
\colhead{Date} &
\colhead{[arcsec]} &
\colhead{Angle [$^\circ$]} &
\colhead{} 
}
\startdata
1997 Mar 19 & 1997.2137 & $ 3.6811\pm0.0040$ & $191.864\pm0.065$ & WFPC2 F1042M spike fit \\
1997 May 18 & 1997.3782 & $ 3.7229\pm0.0047$ & $188.996\pm0.076$ & WFPC2 F1042M spike fit \\
2001 Oct 27 & 2001.8209 & $ 5.1909\pm0.0042$ & $132.600\pm0.049$ & WFPC2 F1042M spike fit \\
2002 May 10 & 2002.3562 & $ 5.4271\pm0.0042$ & $128.119\pm0.047$ & WFPC2 F1042M spike fit \\
2002 Oct 20 & 2002.8012 & $ 5.6285\pm0.0041$ & $124.654\pm0.044$ & WFPC2 F1042M spike fit \\
2003 Apr 18 & 2003.2942 & $ 5.8598\pm0.0043$ & $121.202\pm0.044$ & WFPC2 F1042M spike fit \\
2003 Oct 15 & 2003.7879 & $ 6.0894\pm0.0041$ & $117.913\pm0.041$ & WFPC2 F1042M spike fit \\
2004 Aug 15 & 2004.6224 & $ 6.4675\pm0.0040$ & $113.025\pm0.038$ & WFPC2 F1042M spike fit \\
2005 Apr 20 & 2005.3012 & $ 6.7901\pm0.0042$ & $109.353\pm0.037$ & WFPC2 F1042M spike fit \\
2006 Jan 15 & 2006.0415 & $ 7.1261\pm0.0041$ & $105.823\pm0.034$ & WFPC2 F1042M spike fit \\
2006 Dec 27 & 2006.9883 & $ 7.5551\pm0.0041$ & $101.725\pm0.033$ & WFPC2 F1042M spike fit \\
2008 Jan 03 & 2008.0072 & $ 7.9858\pm0.0041$ & $ 97.585\pm0.031$ & WFPC2 F1042M spike fit \\
\noalign{\smallskip}
2010 Sep 02 & 2010.6697 & $ 9.0212\pm0.0042$ & $ 88.950\pm0.028$ & WFC3 F953N spike fit   \\
2011 Oct 01 & 2011.7484 & $ 9.3939\pm0.0041$ & $ 85.925\pm0.026$ & WFC3 F953N spike fit   \\
2012 Sep 26 & 2012.7392 & $ 9.7126\pm0.0050$ & $ 83.363\pm0.030$ & WFC3 F953N spike fit   \\
2014 Mar 31 & 2014.2455 & $10.1485\pm0.0040$ & $ 79.860\pm0.023$ & WFC3 F953N spike fit   \\
2016 Aug 20 & 2016.6361 & $10.6960\pm0.0040$ & $ 74.630\pm0.022$ & WFC3 F953N spike fit   \\
\noalign{\smallskip}
2010 Sep 02 & 2010.6697 & $ 9.0235\pm0.0040$ & $ 88.954\pm0.027$ & WFC3 F953N PSF fit    \\
2011 Oct 01 & 2011.7484 & $ 9.3973\pm0.0040$ & $ 85.950\pm0.025$ & WFC3 F953N PSF fit    \\
2012 Sep 26 & 2012.7392 & $ 9.7129\pm0.0040$ & $ 83.404\pm0.025$ & WFC3 F953N PSF fit    \\
2014 Mar 31 & 2014.2455 & $10.1424\pm0.0040$ & $ 79.840\pm0.023$ & WFC3 F953N PSF fit    \\
2016 Aug 20 & 2016.6361 & $10.7006\pm0.0040$ & $ 74.660\pm0.022$ & WFC3 F953N PSF fit    \\
\noalign{\smallskip}
2010 Sep 02 & 2010.6697 & $ 9.0224\pm0.0029$ & $ 88.952\pm0.019$ & WFC3 F953N average    \\
2011 Oct 01 & 2011.7484 & $ 9.3956\pm0.0029$ & $ 85.938\pm0.018$ & WFC3 F953N average    \\
2012 Sep 26 & 2012.7392 & $ 9.7128\pm0.0031$ & $ 83.387\pm0.019$ & WFC3 F953N average    \\
2014 Mar 31 & 2014.2455 & $10.1454\pm0.0028$ & $ 79.850\pm0.016$ & WFC3 F953N average    \\
2016 Aug 20 & 2016.6361 & $10.6983\pm0.0028$ & $ 74.645\pm0.016$ & WFC3 F953N average    \\
\enddata
\end{deluxetable}

\begin{deluxetable}{ccccccc}
\tablewidth{0 pt}
\tabletypesize{\footnotesize}
\tablecaption{USNO 26-inch Photographic Astrometric Measurements of Sirius~B 
  Relative to Sirius~A}
\tablehead{
\colhead{Besselian} &
\colhead{Position} &
\colhead{Separation} &
\colhead{} &
\colhead{Besselian} &
\colhead{Position} &
\colhead{Separation} \\
\colhead{Date} &
\colhead{Angle\tablenotemark{a} [$^\circ$]} &
\colhead{[arcsec]} &
\colhead{} &
\colhead{Date} &
\colhead{Angle\tablenotemark{a} [$^\circ$]} &
\colhead{[arcsec]} 
}
\startdata
1970.1331 & $67.63\pm0.10$ & $11.362\pm0.015$ & & 1976.1460 & $56.56\pm0.05$ & $11.214\pm0.019$ \\    
1970.1930 & $67.52\pm1.13$ & $11.315\pm0.230$ & & 1976.1591 & $56.50\pm0.09$ & $11.205\pm0.019$ \\    
1970.1990 & $67.80\pm0.10$ & $11.368\pm0.030$ & & 1976.1840 & $56.53\pm0.07$ & $11.217\pm0.023$ \\    
1970.2371 & $67.33\pm0.41$ & $11.323\pm0.040$ & & 1976.1949 & $56.37\pm0.02$ & $11.193\pm0.004$ \\    
1970.2430 & $67.29\pm0.05$ & $11.320\pm0.024$ & & 1976.2390 & $56.46\pm0.17$ & $11.225\pm0.035$ \\    
1970.2729 & $67.22\pm0.06$ & $11.243\pm0.009$ & & 1976.2610 & $56.26\pm0.00$ & $11.200\pm0.000$ \\    
1970.2920 & $67.46\pm0.03$ & $11.285\pm0.032$ & & 1976.2629 & $56.24\pm0.01$ & $11.178\pm0.006$ \\    
1970.7990 & $66.21\pm0.03$ & $11.290\pm0.005$ & & 1976.9670 & $54.75\pm0.03$ & $11.086\pm0.023$ \\    
1970.8010 & $66.34\pm0.07$ & $11.310\pm0.010$ & & 1977.1290 & $54.55\pm0.03$ & $11.060\pm0.025$ \\    
1970.9520 & $66.05\pm0.04$ & $11.338\pm0.007$ & & 1977.1510 & $54.59\pm0.07$ & $11.065\pm0.031$ \\    
1971.0179 & $65.95\pm0.16$ & $11.385\pm0.027$ & & 1977.1780 & $54.44\pm0.04$ & $11.055\pm0.008$ \\    
1971.2230 & $65.62\pm0.15$ & $11.398\pm0.036$ & & 1977.2410 & $54.21\pm0.00$ & $10.967\pm0.009$ \\    
1971.2720 & $65.49\pm0.10$ & $11.368\pm0.027$ & & 1977.2679 & $54.29\pm0.15$ & $11.026\pm0.055$ \\    
1971.2830 & $65.34\pm0.04$ & $11.325\pm0.014$ & & 1977.9940 & $52.72\pm0.12$ & $10.879\pm0.029$ \\    
1971.8910 & $64.41\pm0.04$ & $11.495\pm0.006$ & & 1978.1470 & $52.65\pm0.07$ & $10.858\pm0.090$ \\    
1971.9050 & $64.29\pm0.04$ & $11.426\pm0.006$ & & 1978.1500 & $52.53\pm0.10$ & $10.828\pm0.039$ \\    
1971.9160 & $64.24\pm0.10$ & $11.413\pm0.016$ & & 1978.2130 & $52.59\pm0.12$ & $10.620\pm0.006$ \\    
1972.1429 & $63.83\pm0.12$ & $11.364\pm0.014$ & & 1979.2310 & $50.18\pm0.12$ & $10.517\pm0.049$ \\    
1972.1479 & $63.88\pm0.10$ & $11.361\pm0.014$ & & 1979.2450 & $50.24\pm0.15$ & $10.568\pm0.036$ \\    
1972.1510 & $63.79\pm0.06$ & $11.365\pm0.015$ & & 1979.8390 & $49.06\pm0.15$ & $10.446\pm0.029$ \\    
1972.1591 & $63.97\pm0.08$ & $11.339\pm0.027$ & & 1979.9100 & $49.04\pm0.21$ & $10.545\pm0.059$ \\    
1973.0710 & $62.19\pm0.10$ & $11.269\pm0.022$ & & 1980.1680 & $48.33\pm0.13$ & $10.327\pm0.052$ \\    
1973.1560 & $62.04\pm0.11$ & $11.485\pm0.028$ & & 1980.2220 & $48.23\pm0.14$ & $10.295\pm0.040$ \\    
1973.7880 & $60.70\pm0.16$ & $11.280\pm0.033$ & & 1980.2390 & $48.08\pm0.11$ & $10.285\pm0.039$ \\    
1973.8430 & $60.82\pm0.06$ & $11.270\pm0.025$ & & 1981.1180 & $46.15\pm0.10$ & $ 9.993\pm0.048$ \\    
1973.9550 & $60.62\pm0.26$ & $11.303\pm0.074$ & & 1981.1230 & $46.20\pm0.14$ & $10.007\pm0.022$ \\    
1974.0430 & $58.56\pm0.09$ & $11.335\pm0.022$ & & 1981.1370 & $46.20\pm0.02$ & $ 9.992\pm0.023$ \\    
1974.2560 & $60.12\pm0.08$ & $11.398\pm0.041$ & & 1981.1560 & $46.10\pm0.09$ & $ 9.987\pm0.058$ \\    
1974.2590 & $60.08\pm0.03$ & $11.348\pm0.015$ & & 1981.1591 & $46.08\pm0.11$ & $ 9.955\pm0.039$ \\    
1974.2650 & $58.46\pm0.00$ & $11.437\pm0.000$ & & 1981.2740 & $45.86\pm0.08$ & $ 9.985\pm0.015$ \\    
1975.2230 & $58.30\pm0.05$ & $11.313\pm0.016$ & & 1981.9860 & $43.94\pm0.09$ & $ 9.704\pm0.018$ \\    
1975.2810 & $58.04\pm0.00$ & $11.288\pm0.000$ & & 1982.2729 & $43.26\pm0.24$ & $ 9.563\pm0.020$ \\    
1975.8440 & $57.02\pm0.07$ & $11.245\pm0.006$ & & 1984.1899 & $37.75\pm0.13$ & $ 8.790\pm0.028$ \\    
\enddata
\tablenotetext{a}{Referred to equator of observation date}
\end{deluxetable}

\begin{deluxetable}{ll}
\tablewidth{0 pt}
\tablecaption{Elements of Relative Visual Orbit of Sirius (J2000)}
\tablehead{
\colhead{Element} &
\colhead{Value} 
}
\startdata
Orbital period, $P$ [yr]                &   50.1284  $\pm$ 0.0043  \\
Semimajor axis, $a$ [arcsec]            &    7.4957  $\pm$ 0.0025  \\
Inclination, $i$ [deg]                  &  136.336   $\pm$ 0.040   \\
Position angle of node, $\Omega$ [deg]  &   45.400   $\pm$ 0.071   \\
Date of periastron passage, $T_0$ [yr]  & 1994.5715  $\pm$ 0.0058  \\
Eccentricity, $e$                       &    0.59142 $\pm$ 0.00037 \\
Longitude of periastron, $\omega$ [deg] &  149.161   $\pm$ 0.075   \\
\enddata
\end{deluxetable}

\begin{deluxetable}{lll}
\tablewidth{0 pt}
\tablecaption{Parallax and Semimajor Axis for Sirius A}
\tablehead{
\colhead{Source} & 
\colhead{Value}  &
\colhead{Reference} 
}
\startdata
\multispan3{\hfil Absolute Parallax, $\pi$ [arcsec] \hfil} \\
\noalign{\vskip-0.25in}\\
Ground-based compilation & $0.3777 \pm 0.0031$ & Gatewood \& Gatewood (1978) \\
\Hipp\/     & $ 0.37921 \pm 0.00158 $ & van Leeuwen (2007)     \\
\noalign{\vskip-0.275in} \\
Weighted mean & $ 0.3789 \pm 0.0014 $ & Adopted       \\
\noalign{\vskip-0.2in} \\
\multispan3{\hfil Semimajor Axis, $a_A$ [arcsec] \hfil} \\
Ground-based compilation & $2.4904 \pm 0.0040$ & Gatewood \& Gatewood (1978) \\
Solution with updated elements & $2.4761 \pm 0.0045$ & Adopted            \\
\enddata
\end{deluxetable}

\begin{deluxetable}{lccc}
\tablewidth{0 pt}
\tabletypesize{\footnotesize}
\tablecaption{Dynamical Masses for Sirius System}
\tablehead{
\colhead{Quantity} & 
\colhead{van den Bos (1960)} &
\colhead{Gatewood \& Gatewood (1978)} &
\colhead{This paper} 
}
\startdata
Total mass, $M_A+M_B$ & $3.20\, M_\odot$ & 
  $3.196 \pm 0.083 \, M_\odot$ & $3.081 \pm 0.034 \, M_\odot$ \\
Mass of Sirius A, $M_A$                 & $2.15\, M_\odot$ & 
  $2.143 \pm 0.056 \, M_\odot$ & $2.063 \pm 0.023 \, M_\odot$ \\
Mass of Sirius B, $M_B$                 & $1.05\, M_\odot$ & 
  $1.053 \pm 0.028 \, M_\odot$ & $1.018 \pm 0.011 \, M_\odot$ \\
\enddata
\end{deluxetable}

\begin{deluxetable}{lllcc}
\tablewidth{0 pt}
\tablecaption{Error Budgets for Sirius System Dynamical Masses}
\tablehead{
\colhead{Quantity} &
\colhead{Value} &
\colhead{Uncertainty} &
\colhead{$\sigma(M_A)$ [$M_\odot$] } &
\colhead{$\sigma(M_B)$ [$M_\odot$] }
}
\startdata
Absolute parallax, $\pi$    & 0.3789  & $\pm$0.0014 arcsec & 0.023  & 0.011  \\
Semimajor axis, $a$         & 7.4957  & $\pm$0.0025 arcsec & 0.0024 & 0.0007 \\
Semimajor axis for A, $a_A$ & 2.4761  & $\pm$0.0045 arcsec & 0.0018 & 0.0018 \\
Period, $P$                 & 50.1284 & $\pm$0.0043 yr     & 0.0004 & 0.0002 \\
\noalign{\vskip0.1in}
Combined mass uncertainty   &         &                    & 0.023  & 0.011 \\
\enddata
\end{deluxetable}

\newcommand{\1}{\phantom{1}}
\newcommand{\n}{\phantom{$-$}}

\begin{deluxetable}{lcccccclccl} 
\tabletypesize{\scriptsize}
\tablewidth{0pt}
\tablecaption{Historical and \HST\/ Astrometry of Sirius B Relative to Sirius A}
\tablehead{
\colhead{Date} & \colhead{PA} & \colhead{Sep.} & \colhead{PA\_cor} &
\colhead{e\_PA} & \colhead{Sep\_cor} & \colhead{e\_Sep} & \colhead{Observer} &
\colhead{Method\tablenotemark{b}} & \colhead{Tel.} & \colhead{Notes\tablenotemark{c}}  \\
\colhead{(BY)} & \colhead{(deg)} & \colhead{(arcsec)} & \colhead{(deg)} &
\colhead{(deg)} & \colhead{(arcsec)} & \colhead{(arcsec)} & \colhead{Code\tablenotemark{a}} &
\colhead{} & \colhead{(m)} & \colhead{}  }
\startdata 
1862.102 & 88.55  &  11.36  &  \n89.369 &$\dots$& 11.3618 &$\dots$ & Bond\_1862a      & M & 0.4 & R1 \\
1862.111 & 85.15  &  10.18  &  \n85.969 & 1.917 & 10.1818 & 0.4467 & Bond\_1862a      & M & 0.4 & M1 \\
1862.127 & 83.00  & \19.85  &  \n83.819 & 1.917 &  9.8518 & 0.4467 & Bond\_1862a      & M & 0.4 &    \\
1862.190 & 84.13  & \19.63  &  \n84.949 & 1.916 &  9.6318 & 0.4466 & Bond\_1862a      & M & 0.4 & M1 \\
1862.239 & 84.15  & \19.94  &  \n84.968 & 1.916 &  9.9418 & 0.4465 & Bond\_1862a      & M & 0.4 & M1 \\
1862.278 & 84.26  &  10.06  &  \n85.078 & 1.916 & 10.0618 & 0.4465 & Bond\_1862a      & M & 0.4 & M1 \\
1862.190 & 85.267 & \18.95  &  \n86.086 & 1.949 &  8.9518 & 0.4615 & Rutherfurd\_1862 & M & 0.3 &    \\
1862.193 & $\dots$&  10.93  & $-$99.000 & 0.000 & 10.9318 & 0.4615 & Rutherfurd\_1862 & M & 0.3 &    \\
1862.215 & 83.0   &  10.4~  &  \n83.818 & 1.785 & 10.4018 & 0.3867 & Chacornac\_1862  & M & 0.8 &    \\
1862.228 & 86.1   &  10.43  &  \n86.918 & 1.785 & 10.4318 & 0.3866 & Chacornac\_1862  & M & 0.8 &    \\
\enddata
\tablenotetext{a}{Observer reference code (see Table 9)}
\tablenotetext{b}{Method code: M = Micrometer, P = Photographic, C = CCD, H =
\HST}
\tablenotetext{c}{Notes codes, explained fully in machine-readable table; R1 =
rejected from solution; M1 = mean of multiple observations; etc.}
\tablecomments{This table is available in its entirety in machine-readable
format through the online journal and VizieR. A portion is shown here for
guidance regarding its form and content.}
\end{deluxetable}

\begin{deluxetable}{lll} 
\tabletypesize{\scriptsize}
\tablewidth{0pt}
\tablecaption{Observer Codes and Literature References for Table~8}
\tablehead{
\colhead{Observer Code} & 
\colhead{BibCode\tablenotemark{a}} & 
\colhead{Full Reference}
}
\startdata 
Aitken\_1896 & 1896PASP....8..314A & Aitken, R.G. 1896, PASP, 8, 314 \\
Aitken\_1914 & 1914PLicO..12....1A & Aitken, R.G. 1914, Publ. Lick Obs., 12, 1 \\
Aitken\_1923 & 1923LicOB..11...58A & Aitken, R.G. 1923, Lick Observatory Bulletin, 11, 58 \\
Aitken\_1926 & 1926PASP...38..131A & Aitken, R.G. 1926, PASP, 38, 131 \\
Aitken\_1927 & 1927LicOB..12..173A & Aitken, R.G. 1927, Lick Observatory Bulletin, 12, 173 \\
Bond\_1862a      & 1862AN.....58...85B & Bond, G.P. 1862, AN, 58, 85	    \\
Chacornac\_1862  & 1862AN.....57..175C & Chacornac, M. 1862, AN, 57, 175     \\
Rutherfurd\_1862 & $\dots$	      & Rutherfurd, L. 1862, AmJSA, 34, 294 \\
\enddata
\tablenotetext{a}{Bibcode in NASA ADS, \tt http://www.adsabs.harvard.edu}
\tablecomments{This table is available in its entirety in machine-readable
format through the online journal and VizieR. A portion is shown here for
guidance regarding its form and content.}
\end{deluxetable}

\end{document}